\documentclass[12pt]{article}
\pdfoutput=1 
     
\usepackage{hyperref}    
\usepackage{amsmath}      
\usepackage{amssymb} 
\usepackage{graphicx}
\usepackage{url}

\usepackage{float,wrapfig,color} 

\allowdisplaybreaks
\def\CR{\nonumber \\} 
\def\eq#1{(\ref{#1})}
\def\s[#1\s]{\begin{align}\begin{split}#1\end{split}\end{align}}
\def\[#1\]{\begin{align}#1\end{align}}

\begin{document}

\begin{titlepage}

\title{
\hfill\parbox{4cm}{ \normalsize YITP-20-34}\\ 
\vspace{1cm} 
Phases of a matrix model \\ with non-pairwise
index contractions}

\author{
Dennis Obster\footnote{dennis.obster@yukawa.kyoto-u.ac.jp} and Naoki Sasakura\footnote{sasakura@yukawa.kyoto-u.ac.jp}
\\
{\small{\it Yukawa Institute for Theoretical Physics, Kyoto University,}}
\\ {\small{\it  Kitashirakawa, Sakyo-ku, Kyoto 606-8502, Japan}}
}

\date{\today}

\maketitle

\begin{abstract}
Recently a matrix model with non-pairwise index contractions has been studied
in the context of the canonical tensor model, 
a tensor model for quantum gravity in the canonical formalism. 
This matrix model also appears in the same form with different ranges of parameters and variables,
when the replica trick is applied to the spherical $p$-spin model ($p=3$) in spin glass theory.
Previous studies of this matrix model suggested
the presence of a continuous phase transition around $R\sim N^2/2$, where $N$ and $R$ 
designate its matrix size $N\times R$.
This relation between $N$ and $R$ intriguingly agrees with a consistency condition of the tensor 
model in the leading order of $N$, suggesting that the tensor model is located near or on the continuous phase 
transition point and therefore its continuum limit is automatically taken in the $N\rightarrow \infty$ limit.
In the previous work, however, the evidence for the phase transition was not 
satisfactory due to the slowdown of the Monte Carlo simulations.
In this work, we provide a new setup for Monte Carlo simulations by integrating out the radial direction
of the matrix. This new strategy considerably improves the efficiency, and allows us
to clearly show the existence of the phase transition.
We also present various characteristics of the phases, 
such as dynamically generated dimensions of configurations, cascade symmetry breaking, and a parameter 
zero limit,
to discuss some implications to the canonical tensor model.
\end{abstract}

\end{titlepage}

\section{Introduction}
\label{sec:introduction}
Quantization of gravity is one of the most challenging fundamental problems in physics, and 
various approaches to this problem have been proposed so far.
These include sophisticated applications
of the renormalization group procedure to general relativity \cite{Reuter:2019byg}
as well as approaches that use discretization of spacetime in the definition of theory, 
for instance the approaches in~\cite{Loll:2019rdj,Rovelli:2014ssa,Surya:2019ndm, Konopka:2006hu},
matrix models~\cite{Wigner,thooft-planar,Brezin:1990rb,Douglas:1989ve,Gross:1989vs},
and tensor models~\cite{Ambjorn:1990ge,Sasakura:1990fs,Godfrey:1990dt,Gurau:2009tw}.
One of the goals of these discretized approaches is to show the emergence of macroscopic spacetime as a continuous manifold,  
with general relativity emerging as  the effective description of dynamics.  
This is still a challenging goal for any of these approaches.

In this paper, we study the dynamics of a matrix model which contains non-pairwise index 
contractions~\cite{Lionni:2019rty,Sasakura:2019hql}. This matrix model has only the $O(N)\times S_R$
symmetry for the index spaces of the matrix variable $\phi_a^i\ (a=1,2,\ldots,N,\ i=1,2,\ldots,R)$, 
where $S_R$ denotes the symmetric group and $O(N)$ the orthogonal group. 
The symmetry is not enough to diagonalize an arbitrary matrix, and therefore this matrix model is not 
solvable by the methods usually employed to solve matrix models~\cite{Wigner,thooft-planar,Brezin:1990rb,Douglas:1989ve,Gross:1989vs}
or rectangular matrix models~\cite{Anderson:1990nw,Anderson:1991ku,Myers:1992dq}.
Our matrix model can also be regarded as a vector model of $R$ vector variables, but our setup is different 
from the exactly solved ones in \cite{Nishigaki:1990sk,DiVecchia:1991vu}.

The background motivation for our matrix model comes from the fact that this model has an 
intimate connection~\cite{Lionni:2019rty,Sasakura:2019hql}
to an exact wave function~\cite{Narain:2014cya} of a tensor model in the Hamilton formalism, 
which we call the canonical tensor model~\cite{Sasakura:2011sq,Sasakura:2012fb}. 
Previously it has been found that this wave function peaks around Lie-group symmetric configurations of the tensor-variable 
of the model~\cite{Obster:2017dhx,Obster:2017pdq}.
This is encouraging towards potential emergence of a spacetime as mentioned above, 
because Lie-group symmetries, such as Lorentz, deSitter, gauge, and so on, are ubiquitous 
in the universe. 
However, it is still difficult to show whether the peaks contain configurations which can be interpreted 
as some sort of spacetime, for instance, in the manner described for a classical treatment in~\cite{Kawano:2018pip}. 
Understanding the properties of the dynamics of our matrix model will potentially provide useful 
insights about the relation between the wave function of the tensor model and spacetime emergence.

It is an intriguing coincidence that a matrix model with the same form has previously appeared in the 
context of spin glasses.  It is obtained, when the replica trick is applied to the spherical $p$-spin
model ($p=3$) \cite{pspin,pedestrians} for spin glasses, where $R$ designates the replica number. 
However, the physics of the spin glass and that of our model will be largely different, because
the parameter and variable regions of interests are different from each other. 
In the spin glass case, the replica number $R$ is taken to the limit $R\rightarrow 0$ as part of its process,
while our interest is rather in the limit $R\sim N^2/2 \rightarrow \infty$, the reason of which 
comes from the consistency of the tensor model, as explained more in Section~\ref{sec:tensor}.  
In addition, the coupling parameter (called $\lambda$ in later sections) of the models has opposite signs,
and the spin glass case has a spherical constraint, $\phi_a^i\phi_a^i=1$, for each $i$. 
Considering these differences, it seems necessary to analyze our model 
independently from the spin glass case. 

In the previous paper \cite{Sasakura:2019hql}, 
Monte Carlo simulations of the model were performed with the usual Metropolis update
method. This has revealed various interesting characteristics of the model.
However, there was an issue which affects the reliability of the Monte Carlo simulations:
For some values of the parameters important to study its properties, 
the iterative updates in the radial direction of the matrix variable were too slow to reach 
thermodynamic equilibriums in a reasonable amount of time.  
For instance, 
it could not be determined with confidence whether the transition is a phase transition
or just a crossover, since the parameters could not be tuned to make the transition more evident.
The major improvement of the present paper is that we integrate out the troublesome radial direction
before doing the numerical calculation, obtaining a model essentially defined on a compact manifold (the hypersphere). 
In addition, the more efficient Hamiltonian Monte Carlo method is employed instead of the more straightforward Metropolis algorithm.
This replacement of the model drastically improves the efficiency of the simulations,
and we have successfully obtained much more evident results than the previous ones. 

We summarize below the properties of the transition and the phases derived from the numerical results:
\begin{itemize}
\item The transition becomes sharper as $N$ is taken larger.  
This implies the transition is a phase transition
in the thermodynamic limit. We have not observed any discrete behavior of observables around the transition point, implying that the transition is continuous.
\item The value of $R$ at the transition point, which we call the critical value $R_c$,
is a little smaller than $(N+1)(N+2)/2$,
that was previously obtained by the perturbative analytic computations in \cite{Lionni:2019rty, Sasakura:2019hql}. 
The critical value $R_c$ is better approximated by $R_c\sim (N+1)(N+2)/2-N+2$
in our numerical results, where the parameters of the model are taken in the range 
$k/\lambda \gtrsim O(10^{-10})$ and $N\lesssim 12$.
\item
It has been shown that 
the Monte Carlo results and the results of the perturbative analytical computations 
in \cite{Lionni:2019rty, Sasakura:2019hql} do not agree with each other near $R\sim R_c$.
The ratios between them on the peaks increase with the decrease of $k/\lambda $, and increase or 
converge\footnote{We could not conclude which one is the right 
behavior from the simulation datas, as we will see later.}
to some $k/\lambda$-dependent limiting values with the increase of $N$. 
Away from $R\sim R_c$, the two approach each other, quickly for $R>R_c$ and gradually for $R<R_c$. 
\item 
It has been reported in \cite{Sasakura:2019hql} that the dimensions of the configurations change 
under the change of $R$ in the vicinity of the transition point $R_c$. This behavior is 
more precisely investigated in this paper. For small $k/\lambda$, 
the dimensions take the smallest values at the transition point,
and take larger values as $R$ is taken further away from $R_c$.
\item
Though we worked with the improved setup explained above,
we still encountered rapid slowdown of iterative updates of Monte Carlo simulations in the parameter
region $R\gtrsim R_c$ and $k/\lambda\lesssim O(10^{-8})$. 
However, the slowdown seemed to be smoothly improved by taking smaller step sizes 
and performing longer simulations. This implies that there is no transition associated 
to the slowdown. Thus, for $R\gtrsim R_c$, the model behaves like a fluid with a viscosity
which continuously grows as $k/\lambda$ decreases. 
\item
For $R\lesssim R_c$,  it has been observed that the dynamics of the model converges in the 
$k/\lambda \rightarrow +0$ limit, in which expectation values of observables and 
the free energy converge to finite values. 
\item
For $R\gtrsim R_c$, it has been observed that the free energy
diverges in the limit $k/\lambda \rightarrow +0$. 
\item
$SO(N)$ symmetry breaking occurs at $R\gtrsim R_c$ due to large $\phi_a^i$. 
This occurs in a cascade manner as $R$ increases:
the breaking of $SO(N)$ occurs first, then $SO(N-1)$, \ldots, and finally $SO(2)$ breaks down. 
\end{itemize}

We also discuss some implications of the numerical results to the tensor model. The most important is 
the coincidence between the location of the transition point and a consistency condition of the tensor model
in the leading order of $N$.
Combining this with the result that the phase transition is continuous, this suggests the possibility that a continuum
theory can be associated to the tensor model. Moreover, the fact that the transition point is where the
dimensions of the configurations quickly decrease towards low values suggests the possibility of 
emergent spacetimes with sensible dimensions in the tensor model.
       
This paper is organized as follows. In Section~\ref{sec:setup}, we explain the matrix model and 
derive the new setup for the numerical simulations, 
which is obtained by integrating out the radial direction of the matrix variable.
In Section~\ref{sec:observables}, observables are introduced. There are roughly two classes of observables, 
one directly related to the matrix model, and the other directly related to our setup.
The two classes are connected by a formula.
In Section~\ref{sec:analytical}, 
we derive some formulas which compute the expectation values of some observables by using
the analytic results obtained previously in \cite{Lionni:2019rty,Sasakura:2019hql}.
In Section~\ref{sec:MonteCarlo}, we comment on our actual Hamiltonian Monte Carlo method 
for the angular variables in our setup.
In Section~\ref{sec:results}, we summarize our results of the Monte Carlo simulations.
In Section~\ref{sec:phasetrans}, we present several pieces of evidence of the phase transition.
In Section~\ref{sec:comparison}, we compare the results of the numerical simulations and 
the analytic perturbative computations, and show that there are differences in the vicinity of the 
transition point, which grow or converge to some $k/\lambda$-dependent values as $N$ increases.
In Section~\ref{sec:k0limit}, the $k/\lambda\rightarrow +0$ limit is discussed. Its behaviour
severely differs in the two phases.
In Section~\ref{sec:geometry}, the geometry of dominant configurations is discussed. 
In particular, the dimensions take minimum values at the transition point. 
In Section~\ref{sec:symmetry}, symmetry breaking in a cascade manner for $R\geq R_c$ is shown. 
In Section~\ref{sec:slowdown}, the slowdown of iterative updates in our simulations is discussed. 
This appears to occur quickly as $k/\lambda$ becomes smaller at $k/\lambda\lesssim O(10^{-8})$
in our simulations, but a quantitative investigation shows that this is a smooth change, implying that there is no 
transition to another phase with slow dynamics.
In Section~\ref{sec:tensor}, the implications of the numerical results to the tensor model are discussed.
In Section~\ref{sec:critical}, the coincidence of the transition point with a consistency condition
of the tensor model is discussed.  
In Section~\ref{sec:expdimensions}, the behavior of the dimensions is explained from the 
symmetry-peak relation argued in \cite{Obster:2017dhx,Obster:2017pdq}.
In Section~\ref{sec:normalizability}, the normalizability of the wave function of the tensor model 
is discussed.
The last section is devoted to a summary and future prospects.

\section{The matrix model and the setup for simulations}
\label{sec:setup}
The matrix model we consider in this paper is defined by the partition function,
\[
Z_{N,R}(\lambda,k):=\int_{\mathbb{R}^{NR}} \prod_{a=1}^{N}\prod_{i=1}^R d\phi_{a}^i
\exp\left(
-\lambda \sum_{i,j=1}^R U_{ij}(\phi)-k \sum_{i=1}^R  U_{ii}(\phi)
\right),
\label{eq:partition}
\] 
where $\phi_a^i\ (a=1,2,\ldots,N,\ i=1,2,\ldots,R)$ denote the matrix variable, 
the integration is over the whole $NR$-dimensional real space,
and the coupling parameters, $k$ and $\lambda$, are assumed to be positive real for the convergence
of the integral as will be explained in more detail below.
Here $U_{ij}(\phi):=(\phi^i_a \phi_a^j)^3$, where the repeated lower indices are assumed to be summed over.
Throughout this paper, repeated lower indices always appear pairwise, and we assume the common convention 
they are summed over, unless otherwise stated.
On the other hand, the upper indices are triply or sixfold contracted in \eq{eq:partition},
and summation over them will always be written explicitly.

The matrix model \eq{eq:partition} has the $O(N)\times S_R$ symmetry, 
where $O(N)$ denotes 
the orthogonal group transformation in the $N$-dimensional vector space of the lower index, and 
$S_R$ denotes the permutation symmetry for the upper index values $\{1,2,\ldots,R\}$. 
The $O(N)\times S_R$ symmetry is generally not enough to diagonalize the matrix $\phi_a^i$, and therefore
the model cannot exactly be solved by the well-known methods often applied to the usual matrix models  
\cite{Wigner,thooft-planar,Brezin:1990rb,Douglas:1989ve,Gross:1989vs}
or the rectangular matrix models \cite{Anderson:1990nw,Anderson:1991ku,Myers:1992dq}.  
Because of the $O(N)$ symmetry, the model can also be regarded as a vector model
 \cite{Nishigaki:1990sk,DiVecchia:1991vu} with the multiplicity of vectors labeled by the upper index.
In fact, our model can be solved in the $N\rightarrow \infty$ limit with finite $R$~\cite{Lionni:2019rty},
as in the vector models \cite{Nishigaki:1990sk,DiVecchia:1991vu} and in the spherical $p$-spin model
\cite{pspin,pedestrians}. 
However, this solution is not so useful, because our major interest is the vicinity of the phase transition point 
with $R\sim N^2/2$, as will be explained later.
 
The first term of the exponent of the matrix model \eq{eq:partition} is positive semi-definite,
since 
\[
\sum_{i,j=1}^R U_{ij}(\phi)=
\sum_{i,j=1}^R (\phi^i_a \phi_a^j)^3=\left(\sum_{i=1}^R \phi_a^i\phi_b^i\phi_c^i\right)
\left(\sum_{j=1}^R \phi_a^j\phi_b^j\phi_c^j\right)
\geq 0.
\label{eq:positiveU}
\]
The equality on the rightmost is actually satisfied by various configurations, 
including straightforward ones like $\phi_a^1=-\phi_a^2,\ldots$. 
Moreover, when $R$ is larger than a certain value, there will be a continuous infinite number of solutions.\footnote{
A simple counting of degrees of freedom implies that the dimension of the solution space of  
$\sum_{i=1}^R \phi_a^i\phi_b^i\phi_c^i=0$ will be given by $NR-N(N+1)(N+2)/6$,
where the former counts the degrees of freedom of $\phi_a^i$ and the latter
the number of independent conditions. 
Therefore, in general for $R> (N+1)(N+2)/6$, the solutions to the equality will exist continuously.}
Therefore, if $k=0$,
it is not obvious whether the integral \eq{eq:partition} is convergent or not. 
On the other hand, if $k>0$, one can immediately see
\[
Z_{N,R}(\lambda,k)<\left( \int_{\mathbb{R}^N} \prod_{a=1}^N d\phi_a e^{-k (\phi_a\phi_a)^3}\right)^R<\infty.
\]
Thus $k>0$ assures the convergence of the integral \eq{eq:partition} for general cases,
while it will be shown later that the $k/\lambda \rightarrow +0$ limit can be taken if $R<R_c$. 

As explained above, the second term in the exponent of \eq{eq:partition}
acts as a regularization of the integral.
A term with the same role existed in the previous studies of the model \cite{Lionni:2019rty,Sasakura:2019hql},
but had a different, namely quadratic, form,  $k\sum_{i=1}^R\phi_a^i\phi_a^i$. 
The main reason for this choice of quadratic form was that then the action (the exponent) had
the standard form used in perturbative computations. 
It is however not necessary to take a quadratic term as a regularization term for the 
perturbative computations\footnote{This was implicitly carried out in~\cite{Lionni:2019rty,Sasakura:2019hql} as well.}, as we will review in Section~\ref{sec:analytical}.
The present choice $\sum_{i=1}^R U_{ii}(\phi)$, which has the same order as the first term, 
is more convenient in the current analysis, 
because then the radial direction of $\phi_a^i$ can be integrated out 
in a straightforward way, as we will perform below. 
Since the radial direction was the main source of the difficulties in the previous
simulations \cite{Sasakura:2019hql}, the present choice will ease the
deadlock of the simulations.   

Now let us divide $\phi_a^i$ into the radial and the angular coordinates,  $\phi_a^i=r \tilde \phi_a^i$, where
$r$ denotes the radial coordinate, and $\tilde \phi_a^i$ denote the angular coordinates.
Putting this reparameterization into \eq{eq:partition} and integrating over $r$, one obtains
\[
Z_{N,R}(\lambda,k)&=\int_{S^{NR-1}} d\tilde \phi\int_0^\infty dr \, r^{NR-1}
\exp\left(
-\left(\lambda \sum_{i,j=1}^R U_{ij}(\tilde \phi)+k \sum_{i=1}^R   
U_{ii}(\tilde \phi)
\right)r^6
\right) \CR
&=\frac{1}{6}\,\Gamma\left(\frac{NR}{6}\right)\int_{S^{NR-1}}d\tilde \phi \left( 
\lambda \sum_{i,j=1}^R U_{ij}(\tilde \phi)+k \sum_{i=1}^R   U_{ii}(\tilde \phi)
\right)^{-\frac{NR}{6}},
\label{eq:zangle}
\]
where $\Gamma(\cdot)$ is the gamma function, and $S^{NR-1}$ denotes the unit $NR-1$ 
dimensional sphere.
We will use \eq{eq:zangle} as the weight of our simulations, where the variables are only the angular ones. 
Our implementation of the Hamiltonian Monte Carlo method for this system
will briefly be explained in Section~\ref{sec:MonteCarlo}.  

Finally, let us comment about the relation between the matrix model \eq{eq:partition} and 
the spherical $p$-spin model for spin glasses, leaving the relation to the canonical tensor model 
for Section~\ref{sec:tensor}.
The partition function of the spherical $p$-spin model for $p=3$ is given by 
\[
Z_{p\hbox{-}{\rm spin}}(P):=\int_{\phi_a\phi_a=1} d\phi\, \exp\left(-P_{abc} \phi_a \phi_b \phi_c\right),
\]
with a random real coupling $P_{abc}$ to simulate a spin glass system. Considering $R$ replicas of the 
same system in the replica trick and simulating the random coupling by a Gaussian distribution 
$e^{-\alpha P_{abc} P_{abc}}$ with positive $\alpha$, one obtains
\[
\int_{\mathbb{R}^{\# P}} \prod_{a,b,c=1 \atop a\leq b \leq c}^N dP_{abc}\,
e^{-\alpha P_{abc}P_{abc} } \left(Z_{p\hbox{-}{\rm spin}}(P)\right)^R={\cal N}
\int_{\phi_a^i\phi_a^i=1} \prod_{i=1}^R d\phi^i
\exp\left(
\frac{1}{4 \alpha} \sum_{i,j=1}^R U_{ij}(\phi)
\right),
\]
where $\cal N$ is an overall coefficient.
The righthand side has a similar form as \eq{eq:partition}, but there are two major differences.
There are the restrictions, $\phi_a^i\phi_a^i=1$ for each $i$, which assure the finiteness of the integration, 
taking the role of the second term in the exponent of \eq{eq:partition}. 
The other difference is that the coefficient of the exponent has the inverse sign compared to \eq{eq:partition}.
This physically means that the dominant configurations will be largely different between the matrix model  \eq{eq:partition} and that of the spherical $p$-spin model.  
In addition, the $R\rightarrow 0$ limit is finally taken as part of the replica trick,
while this is not necessary in the matrix model \eq{eq:partition} itself.
We are rather interested in the dependence on $R$ of the system, especially 
in the regime $R\sim N^2/2$, as we will see later.
In particular, the last relation requires $R\rightarrow \infty$ in the thermodynamic limit $N\rightarrow \infty$,
which is opposite to the spin glass case.

\section{Expectation values of observables}
\label{sec:observables}
For convenience, let us first slightly generalize the definition of the partition function \eq{eq:partition} 
of the matrix model to
\[
Z_{N,R}(\Lambda):=\int_{\mathbb{R}^{NR}} \prod_{a=1}^{N}\prod_{i=1}^R d\phi_{a}^i
\exp\left(
- \sum_{i,j=1}^R \Lambda_{ij} U_{ij}(\phi)
\right).
\label{eq:partitiongen}
\] 
We assume the symmetric matrix coupling $\Lambda$ is taken so that 
the integral is convergent. This includes the original case \eq{eq:partition}
with $\Lambda=\Lambda^{\lambda,k}$, where
\[
\Lambda^{\lambda,k}_{ij}:=\lambda+k \,\delta_{ij}
\label{eq:Lambda}
\]
with positive $\lambda,k$.

Let us introduce
\s[
&z_{N,R}(\Lambda,\beta):=\int_{S^{NR-1}}d\tilde \phi \left( 
\sum_{i,j=1}^R  \Lambda_{ij} U_{ij}(\tilde \phi)
\right)^{-\beta}, \\
&z_{N,R}(\Lambda,\beta,{\cal O}):=\int_{S^{NR-1}}d\tilde \phi \
{\cal O}(\tilde \phi) \left( 
\sum_{i,j=1}^R  \Lambda_{ij} U_{ij}(\tilde \phi)
\right)^{-\beta}, 
\label{eq:defofsmallz}
\s]
where $U_{ij}(\tilde \phi):=(\tilde \phi^i_a \tilde \phi_a^j)^3$, and ${\cal O}(\tilde \phi)$ 
is an arbitrary observable expressed as a function of $\tilde \phi_a^i$. 
As derived in Section~\ref{sec:observables}, by integrating over $r$, 
the partition function \eq{eq:partitiongen} can be expressed by
\[
Z_{N,R}(\Lambda)=\frac{1}{6}\Gamma(\Delta_{NR}) z_{N,R}(\Lambda,
\Delta_{NR}),
\label{eq:ztilde}
\]
where $\Delta_{NR}:=\frac{1}{6}NR$.
From \eq{eq:ztilde}, the expectation value of an observable ${\cal O}(\tilde \phi)$ is given by
\[
\langle O(\tilde \phi) \rangle =\frac{z_{N,R}(\Lambda,\Delta_{NR},{\cal O})}{ z_{N,R}(\Lambda,\Delta_{NR})}.
\]
These are the observables which are directly obtained in our 
Monte Carlo simulations for the angular variables.

There is another kind of observable that is expressed as a function of $\phi_a^i$.
The difference is just the normalization of $\phi_a^i$. Let us introduce a weight 
$[\cdot]$ which counts the multiplicity of $\phi_a^i$ contained in an observable that is assumed to be
a homogeneous function of $\phi_a^i$. 
For example, the weight of $\phi_a^i\phi_a^i$ is given by $[\phi_a^i\phi_a^i]=2$.

Let us consider an observable ${\cal O}(\phi)$ with weight $w$. This can be rewritten as 
${\cal O}(\phi)={\cal O}(\tilde \phi)\, r^w$ by the reparameterization $\phi_a^i=r \tilde \phi_a^i$
with the radial and angular variables.
Then the expectation value is given by
\s[
\langle {\cal O}(\phi) \rangle&=
\frac{1}{Z_{N,R}(\Lambda)}\int_{\mathbb{R}^{NR}} \prod_{a=1}^{N}\prod_{i=1}^R d\phi_{a}^i\,
{\cal O}(\phi)
\exp\left(
-\sum_{i,j=1}^R \Lambda_{ij} U_{ij}(\phi)
\right) \\
&=\frac{1}{Z_{N,R}(\Lambda)}\int_{S^{NR-1}} d\tilde \phi\int_0^\infty dr \, r^{NR-1+w} {\cal O}(\tilde \phi)
\exp\left(
-r^6 \sum_{i,j=1}^R \Lambda_{ij} U_{ij}(\tilde \phi)
\right) \\
&=\frac{\Gamma\left(\Delta_{NR}+w/6\right)z_{N,R}(\Lambda,\Delta_{NR}+w/6,{\cal O})}
{\Gamma(\Delta_{NR})z_{N,R}(\Lambda,\Delta_{NR})} \\
&=
\frac{\Gamma\left(\Delta_{NR}+w/6\right)}{\Gamma(\Delta_{NR})} 
\left\langle {\cal O}(\tilde \phi) \left( \sum_{i,j=1}^R  \Lambda_{ij} U_{ij}(\tilde \phi)
\right)^{-\frac{w}{6}} \right\rangle,
\label{eq:uphicor}
\s]
where we have used the obvious property,
$z_{N,R}(\Lambda,\Delta+w/6,{\cal O})=z_{N,R}(\Lambda,\Delta,{\cal O} \ (\sum_{ij}\Lambda_{ij} U_{ij}(\tilde \phi))^{-w/6})$.
This formula relates the expectation values of the observables of the matrix model \eq{eq:partition} 
expressed by $\phi_a^i$ with those in our Monte Carlo simulations for the angular variables.
This formula is used for deriving some results in Section~\ref{sec:results}.

\section{Analytic computations by a perturbative method}
\label{sec:analytical}
In the previous papers \cite{Lionni:2019rty,Sasakura:2019hql}, the authors introduced a 
function defined by
\[
f_{N,R}(\Lambda) :=\frac{1}{V_{S^{NR-1}}}\int_{S^{NR-1}} d\tilde \phi\, \exp \left( - \sum_{i,j=1}^R \Lambda_{ij}
(\tilde \phi_a^i \tilde \phi_a^j)^3\right),
\label{eq:fnr}
\] 
where $V_{S^{NR-1}}$ designates the volume of the unit sphere, $\int_{S^{NR-1}} d\tilde \phi$,
for the normalization $f_{N,R}(\Lambda=0)=1$. 
This function is related to the partition function \eq{eq:partitiongen} of the matrix model by
\[
Z_{N,R}(\Lambda) =V_{S^{NR-1}} \int_0^\infty dr\, r^{NR-1} f_{N,R}(\Lambda\,r^6).
\label{eq:zbyf}
\]  

A merit of introducing the function \eq{eq:fnr} is that it can obviously be defined for arbitrary complex values 
of $\Lambda_{ij}$ since the integral region in \eq{eq:fnr} is compact, 
meaning that it is an entire function of $\Lambda_{ij}$ \cite{Lionni:2019rty}.
Therefore, if the whole perturbative series expansion in $\Lambda_{ij}$ of this function 
is obtained,  this is convergent for any $\Lambda_{ij} \neq \infty$, and hence determines the function
completely in the whole complex region of $\Lambda_{ij}$. 
This means that, in principle, the dynamics of the matrix model
\eq{eq:partition} can be determined as precisely as one can by improving the perturbative series expansion 
of $f_{N,R}(\Lambda)$. 
 This is in contrast with the partition function $Z_{N,R}(\lambda,k)$,
which is singular at $\lambda=0$ or $k=0$ and merely an asymptotic perturbative series
expansion of it in $\lambda$ or $k$ can be obtained.

The perturbative computations of $f_{N,R}(\Lambda)$ using Feynman diagrams have 
been performed in \cite{Lionni:2019rty,Sasakura:2019hql}. This can be done by mapping
the integrals $\int_{S^{NR-1}} d\tilde \phi\, \tilde \phi_{a_1}^{i_1}\tilde \phi_{a_2}^{i_2}
\cdots \tilde \phi_{a_n}^{i_n}$,
which appear in the expansion of the integrand in \eq{eq:fnr}, 
to the standard computations using Wick contractions.  
The final result derived in the leading order is given by
\[ 
f^{leading}_{N,R}(\Lambda)=\prod_{e_{\Lambda}} h_{N,R}(e_{\Lambda}),
\label{eq:fleadinggen}
\]
where the product is over the eigenvalues of the matrix $\Lambda_{ij}$ with degeneracies taken
into account, and 
\[
h_{N,R}(t):=(1+12 \gamma_3 t)^{-\frac{N(N+4)(N-1)}{12}} \, (1+6 (N+4) \gamma_3 t)^{-\frac{N}{2}} 
\]
with 
\[
\gamma_3:=\frac{\Gamma\left(\frac{NR}{2}\right)}{8\, \Gamma\left( \frac{NR}{2}+3\right)}.
\]
In the case with $\Lambda^{\lambda,k}_{ij}=\lambda+k\,\delta_{ij}$, the eigenvalues are $k+\lambda R$ for the 
eigenvector $(1,1,\ldots,1)$ and $k$ for all the other vectors transverse to that. Therefore
\[
f^{leading}_{N,R}(\Lambda^{\lambda,k} t)=h_{N,R}\left((k+\lambda R)\,t\right) 
\left(h_{N,R}\left(k\,t\right)\right)^{R-1}.
\label{eq:fleading}
\]

To obtain formulas for expectation values of observables, let us introduce
\[
g_{N,R}(\Lambda,\beta):=\int_0^\infty dt\, t^{\beta-1} f_{N,R}(\Lambda t).
\label{eq:gwithf}
\] 
From \eq{eq:fnr}, one finds
\[
g_{N,R}(\Lambda,\beta)=\frac{\Gamma(\beta)}{V_{S^{NR-1}}} \int_{S^{NR-1}} d\tilde \phi 
\left( \sum_{i,,j=1}^R \Lambda_{ij}
(\tilde \phi_a^i \tilde \phi_a^j)^3\right)^{-\beta}.
\]
Therefore, it has a relation with \eq{eq:defofsmallz} as
\[
g_{N,R}(\Lambda,\beta)=\frac{\Gamma(\beta)}{V_{S^{NR-1}}}  z_{N,R}(\Lambda,\beta). 
\]
Then, by comparing with the results in Section~\ref{sec:observables}, 
the correlation functions of $U_{ij}(\tilde \phi):=(\tilde \phi_a^i \tilde \phi_a^j)^3$ for the angular 
variables can be expressed as
\s[
\langle U_{i_1j_1}(\tilde \phi)  \cdots  U_{i_Mj_M}(\tilde \phi)\rangle&:= \frac{
\int_{S^{NR-1}} d\tilde \phi \,  U_{i_1j_1}(\tilde \phi) \cdots  U_{i_Mj_M}(\tilde \phi)
\left( \sum_{i,,j=1}^R \Lambda_{ij}
(\tilde \phi_a^i \tilde \phi_a^j)^3\right)^{-\Delta_{NR}}}{\int_{S^{NR-1}} d\tilde \phi 
\left( \sum_{i,,j=1}^R \Lambda_{ij}
(\tilde \phi_a^i \tilde \phi_a^j)^3\right)^{-\Delta_{NR}}} \\
&=\frac{(-1)^M}{g_{N,R}(\Lambda,\Delta_{NR})} \frac{\partial}{\partial \Lambda_{i_1j_1}}\cdots \frac{\partial}{\partial \Lambda_{i_Mj_M}}
g_{N,R}(\Lambda,\Delta_{NR}-M).
\label{eq:obsgphitilde}
\s] 
Therefore, by combining with \eq{eq:fleadinggen} (or \eq{eq:fleading}) and \eq{eq:gwithf} and 
numerically integrating over $t$, one can compute the correlation functions of  $U_{ij}(\tilde \phi)$ 
in the leading order of the analytic perturbative computation.

Let us next consider the correlation functions of $U_{ij}(\phi)$ for the variable $\phi_a^i$.
We can use the formula \eq{eq:uphicor}, where each of $U_{ij}(\phi)$ has weight $w=6$. 
We obtain
\s[
\langle U_{i_1j_1}(\phi)  \cdots  U_{i_Mj_M}(\phi)\rangle
&=
\frac{\Gamma\left(\Delta_{NR}+M\right)}{\Gamma(\Delta_{NR})} 
\left\langle U_{i_1j_1}(\tilde \phi)  \cdots  U_{i_Mj_M}(\tilde \phi) 
\left( \sum_{i,j=1}^R  \Lambda_{ij} U_{ij}(\tilde \phi)
\right)^{-M} \right\rangle \\
&=\frac{(-1)^M}{g_{N,R}(\Lambda,\Delta_{NR})}
\frac{\partial}{\partial \Lambda_{i_1j_1}}\cdots \frac{\partial}{\partial \Lambda_{i_Mj_M}}
g_{N,R}(\Lambda,\Delta_{NR}).
\label{eq:obsgphi}
\s] 
This also gives the correlation functions of $U_{ij}(\phi)$ in the leading order from the 
analytic perturbative method.

Later we consider an observable given 
by $U_d(\phi):=\sum_{i=1}^R U_{ii}(\phi)=\sum_{i=1}^R (\phi_a^i\phi_a^i)^3$. 
In our actual case with $\Lambda=\Lambda^{\lambda,k}$ given in \eq{eq:Lambda}, 
the derivatives in \eq{eq:obsgphitilde} and \eq{eq:obsgphi} for the observable can be 
performed by $\frac{\partial}{\partial k}$.
Therefore the correlation functions are given by
\s[
\left\langle (U_d(\tilde \phi))^M \right\rangle&=\frac{(-1)^M}{g_{N,R}(\Lambda^{\lambda,k}, \Delta_{NR})}
\frac{\partial^M}{\partial k^M} g_{N,R}(\Lambda^{\lambda,k}, \Delta_{NR}-M), 
\\
\left\langle (U_d(\phi))^M \right\rangle&=\frac{(-1)^M}{g_{N,R}(\Lambda^{\lambda,k}, \Delta_{NR})}
\frac{\partial^M}{\partial k^M} g_{N,R}(\Lambda^{\lambda,k}, \Delta_{NR}).
\label{eq:od}
\s]
This formula is used when we compare the numerical results with the analytical ones in Section~\ref{sec:comparison}.

\section{Hamiltonian Monte Carlo method for angular variables}
\label{sec:MonteCarlo}
In this paper, we use Hamiltonian Monte Carlo method \cite{HMC} for the numerical simulations. 
This method upgrades the configuration space of some integral 
to a phase space by introducing conjugate variables, 
and creates a Hamilton system and (locally) solves the equations of motion
in order to find new, more remote, candidates for the Metropolis update.
This process is called leapfrog, which consists of a sequence of discrete jumps from one 
phase space location to another. While it is enough for presenting update candidates 
to approximately solve classical equation of motion, 
the time reversal symmetry and the conservation of phase space 
volume must be exactly satisfied under the discrete jumps for correct sampling of configurations.  
For a flat configuration space, these conditions are easily satisfied by alternately sequencing
the following two processes:
\s[
\hbox{(i)  }&\delta q_i= \epsilon\, p_i,\  \delta p_i=0, \\
\hbox{(ii) }&\delta q_i=0,\ \delta p_i= -\epsilon \frac{\partial V(q)}{\partial q_i},
\s]    
where $(q_i,p_i)$ designate phase space variables indexed by $i$, $\epsilon$ is the size of one jump, and $V(q)$ is the Gibbs potential for a weight $\exp(-V(q))$. 
Observe that (i) is a free motion in a flat space, and only (ii) takes effects from $V(q)$.
Each of the two jumps obviously satisfies the conservation of 
the phase space volume,
$\det
|\partial (q_i+\delta q_i), \partial (p_i+\delta p_i)/\partial q_j,\partial p_j|=1,
$
due to the fact that $q_i$ and $p_i$ do not jump simultaneously.
The time reversal symmetry is also satisfied, since $(q_i+\delta q_i,p_i+\delta p_i) \rightarrow (q_i,p_i)$ 
when $\epsilon$ is replaced with $-\epsilon$. 
 
When the configuration space $q_i$ is constrained to a non-flat sub-manifold embedded in a flat space, 
a free motion corresponding to (i) is generally a simultaneous jump of $p_i$ and $q_i$,
since the tangent space of the sub-manifold containing $p_i$ changes along $q_i$.
In such a case, finding an appropriate jump corresponding to (i) satisfying the two necessary 
conditions above is generally a difficult problem. An obvious solution to an appropriate jump is to exactly solve 
the classical equation of the free (geodesic) motion on the sub-manifold \cite{sphere}. This is possible when
a sub-manifold is simple enough to allow us to obtain such exact solutions. 
In our case, the embedded manifold is a unit hypersphere, which gives the constraints, $\sum_i q_i^2=1$ and 
$\sum_{i} q_i p_i=0$, and the jump describing the exact free (geodesic) motion on the sphere is given by 
\[
\hbox{(i') }
\left( 
\begin{matrix}
q'_i \\
p'_i
\end{matrix}
\right)
=\left(
\begin{matrix}
\cos \theta & \frac{\sin \theta}{|p|} \\
-|p| \sin \theta & \cos \theta
\end{matrix}
\right)
\left( 
\begin{matrix}
q_i \\
p_i
\end{matrix}
\right),
\label{eq:rotation}
\] 
where $|p|=\sqrt{\sum_{i}p_i^2}$ and $\theta=\epsilon\, |p|$. 
The second jump (ii) does not contain a jump in $q_i$, therefore 
there are no difficult issues, and it can just be replaced by 
\[
\hbox{(ii') }&\delta q_i=0,\ \delta p_i= -\epsilon \frac{\partial V(q)}{\partial q_i}+\epsilon q_i 
\sum_j q_j\frac{\partial V(q)}{\partial q_j},
\]
where the additional term takes into account the constraint $\sum_i q_i p_i=0$.

In our present case \eq{eq:zangle}, 
the coordinates $\tilde \phi_a^i$ are constrained on a unit sphere $\sum_{i=1}^R \tilde \phi_a^i \tilde \phi_a^i=1$, 
and we employ these jumps (i') and (ii').  
The potential energy can be read from \eq{eq:zangle} as 
\[
V(\tilde \phi_a^i)=\Delta_{NR} \log \left( \sum_{i,j=1}^R \Lambda_{ij} U_{ij}(\tilde \phi) \right)
\] 
with $\Lambda=\Lambda^{\lambda,k}$.

\section{Results of Monte Carlo simulations} 
\label{sec:results}
In this section, we summarize the results of our Hamiltonian Monte Carlo simulations from several view points. 
Since the overall factor of the exponent of \eq{eq:partition} can be absorbed in the rescaling
of $\phi_a^i$, we set $\lambda=1$ in all the simulations, leaving $N$, $R$, and $k$ as variable parameters.
Errors were estimated by the Jackknife method described for example in \cite{Hanada:2018fnp}.
We took the leapfrog numbers to be about 1000-10000, depending on the hardness of the simulations 
explained in Section~\ref{sec:slowdown}, and the step sizes were tuned so that the acceptance rates were about 
80-99 percent, which were a little higher than the commonly taken ones because of the reason 
explained in Section~\ref{sec:slowdown}. 
Parallel tempering \cite{parallel} was also used in some of the computations to take some datas 
which systematically study $k$-dependencies. 
However, as will be explained more in Section~\ref{sec:slowdown}, parallel tempering did not 
seem to essentially affect the expectation values computed.

In the following subsections, we show the results of the simulations of the expectation values of 
various observables depending on the purposes. 
The observables are taken to be invariant under the $O(N)\times S_R$ symmetry.

\subsection{Phase transition point}
\label{sec:phasetrans}
There are various observables which can be used to study the location of the phase transition. We will present 
one example for $\tilde \phi_a^i$ and another for $\phi_a^i$.

The observable we first consider is 
\[
{\cal O}_1:=N \sum_{i,j=1\atop i\neq j}^R (\tilde \phi_a^i \tilde \phi_a^j)^2. 
\]
An important reason for considering this observable is that this has the natural normalization
factor $N$. This factor is determined by the uncorrelated case,
in which each of $\tilde \phi_a^i$ is regarded as an equally independent variable.
More precisely, the uncorrelated case corresponds to
$\langle \tilde \phi_a^i  \tilde \phi_b^j \rangle_{uncorrelated}\sim \delta_{ab}\delta^{ij}/(RN)$
up to sub-leading corrections in $N$ and $R$ by taking into account the constraint, 
$\sum_{i=1}^R \tilde \phi_a^i  \tilde \phi_a^i=1$.
Under this assumption,
\s[
\langle {\cal O}_1 \rangle_{uncorrelated}
&=N  \sum_{i,j=1\atop i\neq j}^R 
\langle \tilde \phi_a^i \tilde \phi_a^j \tilde \phi_b^i \tilde \phi_b^j\rangle_{uncorrelated}\\
&\sim N \sum_{i,j=1\atop i\neq j}^R  \langle \tilde \phi_a^i \tilde \phi_b^i \rangle_{uncorrelated} 
\langle \tilde \phi_a^j \tilde \phi_b^j\rangle_{uncorrelated} \\
&\sim 1, 
\s]
where we have ignored sub-leading corrections in $N$ and $R$. 

\begin{figure}[]
\begin{center}
\hfil
\includegraphics[clip,width=7.0cm]{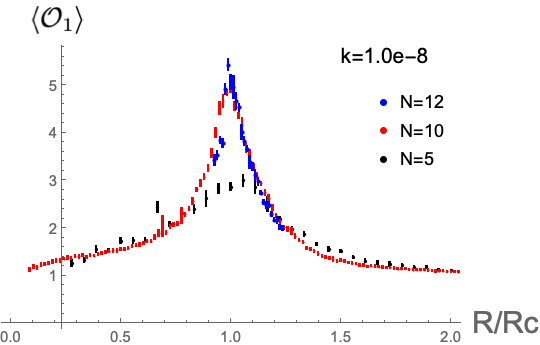}
\hfill
\includegraphics[clip,width=7.0cm]{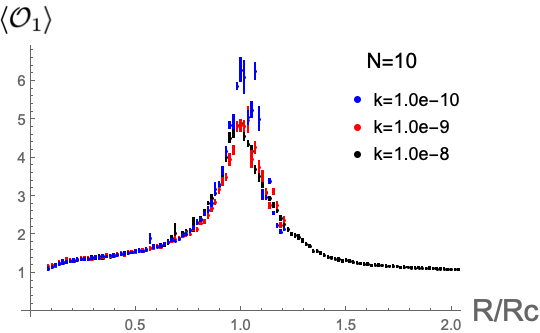}
\hfil
\caption{The results of the Monte Carlo simulations for $\langle {\cal O}_1 \rangle$. 
The horizontal axes are $R/R_c$, where $R_c=(N+1)(N+2)/2-N+2$. }
\label{fig:o1}
\end{center}
\end{figure}

\begin{figure}[]
\begin{center}
\includegraphics[clip,width=10.0cm]{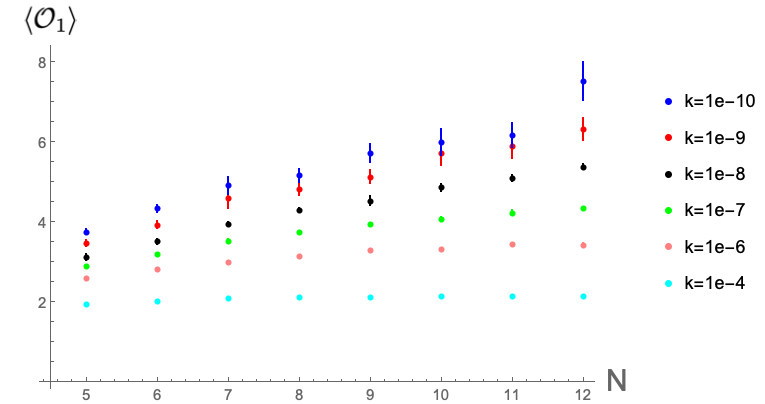}
\caption{$\langle {\cal O}_1 \rangle$ from the simulations are plotted against $N$ with $R=(N+1)(N+2)/2-N+2$.}
\label{fig:peako1}
\end{center}
\end{figure}

Figure~\ref{fig:o1} shows the results of the Monte Carlo simulations for $\langle {\cal O}_1 \rangle$. 
The normalization factor $R_c$ for the horizontal axes is chosen as $R_c=(N+1)(N+2)/2-N+2$. 
The perturbative computations in the leading order predict the transition point to be at 
$R_c=(N+1)(N+2)/2$ \cite{Lionni:2019rty,Sasakura:2019hql}.
However, for the datas shown in the left figure for $k=10^{-8}$, 
it is better to take $R_c=(N+1)(N+2)/2-N+2$ to locate all the peaks near $R/R_c=1$. 
The values of  $\langle {\cal O}_1 \rangle$ approach $1$ as $R$ takes more distant values from $R_c$, 
implying that the correlations become more independent there.
On the other hand, the values of $\langle {\cal O}_1 \rangle$ at the peaks 
become larger for larger $N$. This can be checked more clearly in Figure~\ref{fig:peako1}. 
This means that the correlation becomes larger at the transition point for larger $N$, 
which is a typical signature of a continuous phase transition.

The right picture of Fig.~\ref{fig:o1} shows the dependence of $\langle {\cal O}_1 \rangle$ on $k$ for $N=10$. 
The dependence on $k$ seems little for $R\lesssim R_c$, as we will discuss more of this aspect 
in Section~\ref{sec:k0limit}. 
On the other hand, at $R\gtrsim R_c$, $\langle {\cal O}_1 \rangle$ seems to become larger 
as $k$ becomes smaller, slightly shifting the locations of the peaks to the right. 
This implies that the correlations become larger for smaller $k$ and the critical value $R_c$ 
depends not only on $N$ but also on $k$ as well.   
The last statement implies that what we have taken as $R_c$ above cannot be considered 
to be a correct expression valid for the general values of the parameters, 
but can at most be considered to be an approximate expression 
valid for our parameter range $N\lesssim 12$ and $10^{-10}\lesssim k\lesssim 10^{-8}$.   

\begin{figure}[]
\begin{center}
\hfil
\includegraphics[clip,width=7.0cm]{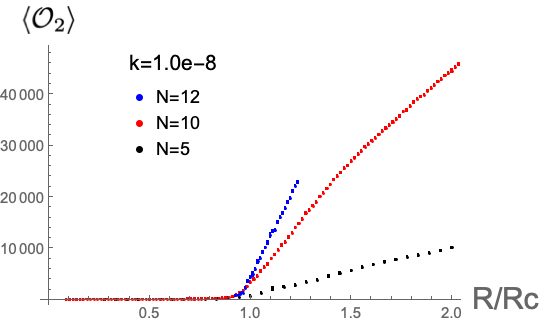}
\hfill
\includegraphics[clip,width=7.0cm]{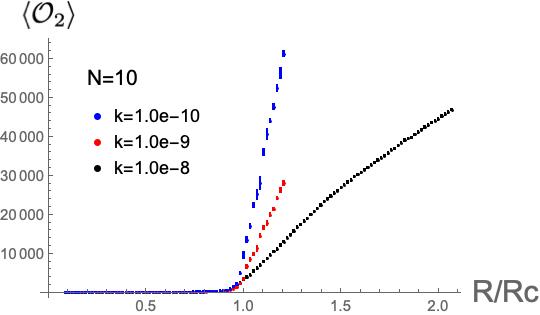}
\hfil
\caption{The results of the Monte Carlo simulations for $\langle {\cal O}_2 \rangle$. 
The horizontal axes are $R/R_c$, where $R_c=(N+1)(N+2)/2-N+2$. }
\label{fig:o2}
\end{center}
\end{figure}

Let us next turn to the observable, 
\[
{\cal O}_2:=\sum_{i=1}^R \phi_a^i \phi_a^i=r^2. 
\]
From the formula \eq{eq:uphicor}, by setting the weight $w=2$ and noting ${\cal O}_2(\tilde \phi)=1$ identically, we obtain
\[
\langle r^2 \rangle = \frac{\Gamma\left(\Delta_{NR}+1/3\right)}{\Gamma(\Delta_{NR})} 
\left\langle \left( \sum_{i,j=1}^R  \Lambda_{ij} U_{ij}(\tilde \phi)
\right)^{-\frac{1}{3}} \right\rangle
\]
with $\Lambda=\Lambda^{\lambda,k}$. 
The results of the simulations for $\langle r^2 \rangle$ are plotted in Figure~\ref{fig:o2}. The figures 
clearly show that the two phases are characterized by $\langle r^2 \rangle \sim 0$ for $R<R_c$ and 
$\langle r^2 \rangle >0$ for $R>R_c$, respectively. The transition becomes sharper as $N$ becomes larger
or $k$ becomes smaller. $\langle r^2 \rangle$ changes continuously at $R\sim R_c$,
supporting the claim that the transition is continuous. 

\subsection{Comparison with the perturbative computation}
\label{sec:comparison}
In this section, we compare the results of the simulations with the analytic perturbative computation
in the leading order, which was reviewed in Section~\ref{sec:analytical}. 
In particular, we see that the analytic computation does not explain the peaks of the correlations of 
$\tilde \phi_a^i$, which was shown in Section~\ref{sec:phasetrans}. 
We find clear deviations between them around the phase transition 
point $R\sim R_c$, while they converge as $R$ takes distant values from $R_c$. 

To see this we consider the observables, $U_d(\tilde \phi)
:=\sum_{i=1}^R U_{ii}(\tilde \phi)$ and $U_d(\phi):=\sum_{i=1}^R U_{ii}(\phi)$,
whose formulas of the analytic computation are given in \eq{eq:od}.
The explicit values are obtained by performing the numerical integration of  
\eq{eq:gwithf} with \eq{eq:fleading} contained in \eq{eq:od} for $M=1$.
On the other hand, we compare these with 
$\langle U_d(\tilde \phi)\rangle$ and 
\[
\langle U_d(\phi)\rangle=\Delta_{NR}\left\langle \frac{U_d(\tilde\phi)}{
\sum_{i,j=1}^R \Lambda^{\lambda,k}_{ij} U_{ij} (\tilde \phi) }
\right\rangle
\label{eq:odphiformula}
\]
from the simulations, where we have used \eq{eq:uphicor} for $w=6$. 

\begin{figure}[]
\begin{center}
\hfil
\includegraphics[clip,width=7.0cm]{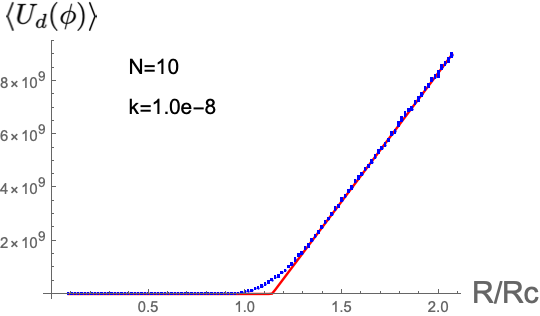}
\hfill
\includegraphics[clip,width=7.0cm]{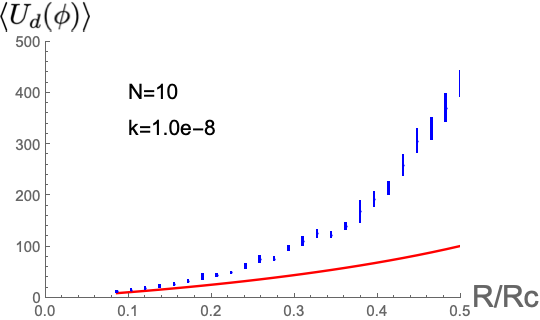}
\hfil
\caption{The comparison between the results of the Monte Carlo simulation
and the analytic perturbative computation of $\langle U_d(\phi) \rangle$ for 
$N=10$ and $k=10^{-8}$. The blue dots with error bars are the Monte Carlo results,
and the red lines are the analytic results.
The horizontal axes are $R/R_c$, where $R_c=(N+1)(N+2)/2-N+2$. The right figure 
magnifies the region $R/R_c<0.5$ in the left figure.}
\label{fig:od}
\end{center}
\end{figure}

Figure~\ref{fig:od} shows the comparison between the Monte Carlo results and the analytic
computations. There exist systematic deviations in the vicinity of $R=R_c$, as was previously reported in
\cite{Sasakura:2019hql}. For $R>R_c$, they quickly converge 
as $R$ leaves $R_c$. For $R<R_c$, they slowly converge as $R$ becomes smaller.

\begin{figure}[]
\begin{center}
\includegraphics[clip,width=8.2cm]{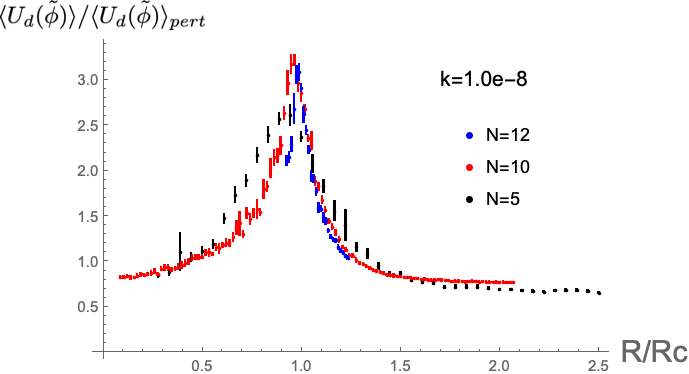}
\includegraphics[clip,width=8.2cm]{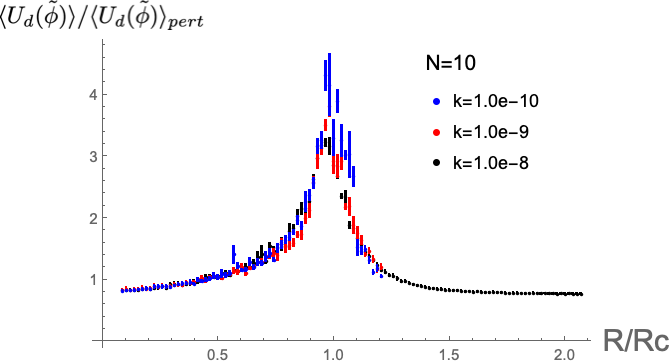}
\caption{The ratio  $\langle U_d(\tilde \phi) \rangle/\langle U_d(\tilde \phi) \rangle_{pert}$ 
between the Monte Carlo and the perturbative analytic results.  
The horizontal axes are $R/R_c$, where $R_c=(N+1)(N+2)/2-N+2$. }
\label{fig:udt}
\end{center}
\end{figure}

\begin{figure}[]
\begin{center}
\includegraphics[clip,width=10cm]{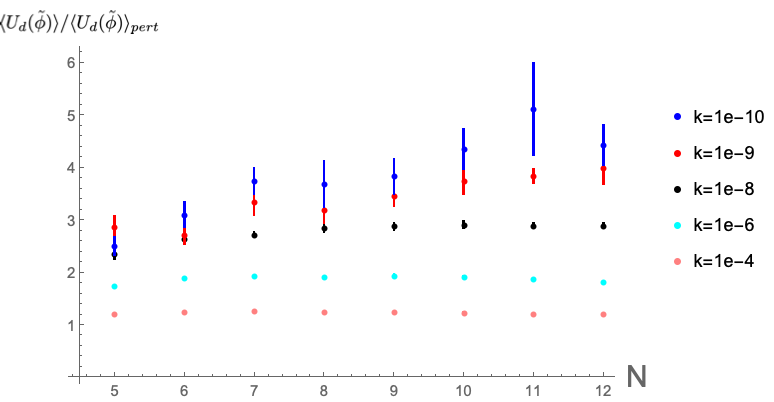}
\caption{The ratio  $\langle U_d(\tilde \phi) \rangle/\langle U_d(\tilde \phi) \rangle_{pert}$ 
from the simulations are plotted against $N$ with $R=(N+1)(N+2)/2-N+2$. }
\label{fig:peakudt}
\end{center}
\end{figure}

One can see similar deviations for the $\langle U_d (\tilde \phi)\rangle$. Figure~\ref{fig:udt} plots the ratio 
$\langle U_d (\tilde \phi)\rangle/\langle U_d (\tilde \phi)\rangle_{pert}$ between the 
Monte Carlo results and the perturbative analytic computations.
Indeed the ratio deviates from 1 in the vicinity of the transition point.  
The deviations 
at the peaks become larger as $k$ becomes smaller. 
On the other hand, as shown in Figure~\ref{fig:peakudt},
it seems that the deviations increase with $N$ for $k<10^{-8}$, but this is not clear for $k\geq 10^{-8}$.
We cannot rule out the possibility that they actually converge in the large $N$ limit to some 
values which increase with the decrease of  $k$. 

From the comparisons above, we conclude that the perturbative analytic computation
in the leading order does not correctly reproduce the behavior of the matrix model
in the vicinity of the transition point.   
As was previously performed in \cite{Sasakura:2019hql}, the situation does not essentially change,
even if we take into account the next leading order corrections to the analytic computation. 

\subsection{$k/\lambda \rightarrow +0$ limit}
\label{sec:k0limit}
In this subsection, we focus on the $k/\lambda \rightarrow +0$ limit of the matrix model \eq{eq:partition}.
There are a few reasons to study this. One is the characterization of the phases separated at $R=R_c$. 
We find different limits for each phase at $R>R_c$ and $R<R_c$. 
Another is its relevance to the tensor model. The behavior determines
whether the wave function is normalizable or not. This will be discussed in Section~\ref{sec:tensor}.

\begin{figure}[]
\begin{center}
\includegraphics[clip,width=7.0cm]{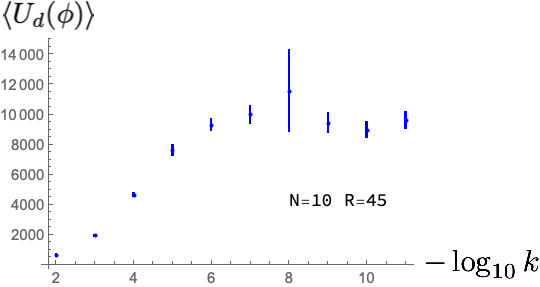}
\caption{The values of $\langle U_d(\phi) \rangle$ from the simulations are plotted against
$-\log_{10} k$ for $N=10,\ R=45$, which belongs to the region $R<R_c$.
The reason for a slightly larger error for $k=10^{-8}$ data point may come from 
the trapping in the narrow region explained in Section~\ref{sec:slowdown}. }
\label{fig:udlow}
\end{center}
\end{figure}

Firstly, let us show that the limit $k/\lambda \rightarrow +0$ converges in the phase $R<R_c$. 
This can be seen by looking at the behavior of expectation values of observables.
Figure~\ref{fig:udlow} shows the result of the simulation
about the behavior of $\langle U_d (\phi)\rangle$ 
in \eq{eq:odphiformula} against $k$ for a case with $R<R_c$. 
As can be seen in the figure, the expectation value approaches a constant value in the $k\rightarrow 
+0$ limit.  
In fact, similar convergence can be observed also for other observables in other cases with $R<R_c$.   

Let us discuss the consequence of this behavior to
 the free energy defined by $F_{N,R}(\lambda,k):=-\log Z_{N,R}(\lambda,k)$. 
By taking the derivative of \eq{eq:partition} with respect to $k$, we obtain
\[
\frac{\partial}{\partial k} F_{N,R}(\lambda,k)=
\langle U_d(\phi)\rangle.
\] 
Therefore, as a function of $k$, $F_{N,R}(\lambda,k)$ 
can be determined by studying the $k$-dependence of $\langle U_d(\phi)\rangle$ and 
performing integration:
\[
F_{N,R}(\lambda,k_1)=\int_{k_0}^{k_1} dk\, \langle U_d(\phi)\rangle+F_{N,R}(\lambda,k_0).
\label{eq:freebyk} 
\]
In particular, $\lim_{k\rightarrow +0} \langle U_d(\phi)\rangle$
will determine $\lim_{k\rightarrow +0} F_{N,R}(\lambda,k)$.

Below let us discuss the behavior of the free energy in $k/\lambda\rightarrow +0$. 
By performing the rescaling of the variable
 $\phi_a^i\rightarrow \lambda^{-1/6} \phi_a^i$ in \eq{eq:partition}, one obtains
 \[
 F_{N,R}(\lambda,k)=F_{N,R}(1,k/\lambda)+\frac{NR}{6}\log \lambda.
\label{eq:freelam}
 \]
In the region $R<R_c$, there is a finite limit of $\lim_{k\rightarrow +0} \langle U_d(\phi)\rangle$
as shown above.
Considering \eq{eq:freebyk} and \eq{eq:freelam}, the behavior of the free energy is obtained as 
\[
F_{N,R}(\lambda,k)=U_d^0 \frac{k}{\lambda} +p_{N,R}(k/\lambda)+\frac{NR}{6} \log \lambda \ \ \hbox{ for } 
k/\lambda\sim +0 \hbox{ and } R<R_c,
\label{eq:freelow}
\] 
where $U_d^0:=\lim_{k\rightarrow +0} \langle U_d(\phi)\rangle_{\lambda=1}$, and 
$p_{N,R}(k/\lambda)$ is smaller than $k/\lambda$ in order and has a finite limit $p_{N,R}(+0)$. 
We comment that this finiteness was proven analytically for $R=2$ and 
any $N$ previously in \cite{Sasakura:2019hql}\footnote{See an appendix of the reference.}.
This finiteness of $F_{N,R}(\lambda,k)$ in the $k\rightarrow +0$ limit
is non-trivial, as discussed in Section~\ref{sec:setup}.

\begin{figure}[]
\begin{center}
\includegraphics[clip,width=7.0cm]{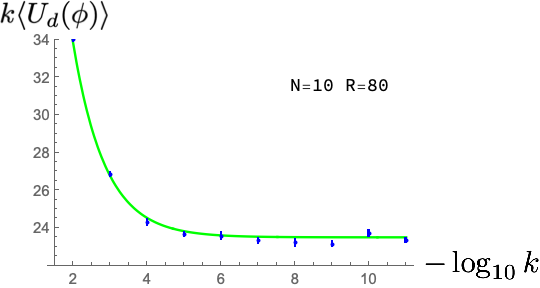}
\caption{$k \langle U_d(\phi) \rangle$ from the simulations is plotted against
$-\log_{10} k$ for $N=10,\ R=80$, which is a case of $R>R_c$.
The data points can be fitted very well with $k \langle U_d(\phi) \rangle \simeq 23.3+ 107 \sqrt{k}$.}
\label{fig:udkdep}
\end{center}
\end{figure}

On the other hand, for $R>R_c$, the simulations show that $\langle U_d (\phi)\rangle$ diverges in the 
$k\rightarrow +0$ limit. An interesting matter is that, instead,  
$k \langle U_d (\phi)\rangle$ converges in the $k\rightarrow  +0$ limit, as can be seen 
from Figure~\ref{fig:udkdep}.
This implies that, from \eq{eq:freebyk}, $F_{N,R}(\lambda,k)$ logarithmically diverges 
in the limit $k/\lambda \rightarrow +0$.

Let us discuss this divergence of the free energy in more detail. As we have seen in Figure~\ref{fig:udkdep}, 
if we take $k$ small enough, $k \langle U_d(\phi) \rangle$ can be regarded as
its limiting value, $\lim_{k\rightarrow +0} k\, \langle U_d(\phi) \rangle$.
By assuming this for the $N=10$ data in the large $R$ region in the left figure of Figure~\ref{fig:od}, and 
fitting a linear function of $R$ for the data in the region $R>1.4\cdot R_c$, one obtains,
\[
 k\, \langle U_d(\phi) \rangle |_{k=10^{-8}}
\simeq 1.66 \cdot ( R-65.9).
\] 
This curiously agrees very well with what can be obtained by putting $N=10$ to a hypothetical expression
for the righthand side,
\[
\frac{N}{6} \left( R-\frac{(N+1)(N+2)}{2}\right)=\frac{NR}{6}-\frac{\# P}{2}
\label{eq:hypothesis}
\]
where $\#P:=N(N+1)(N+2)/6$ is the number of independent components of 
a symmetric three-index tensor $P_{abc}$. We have performed similar analyses for $N=5,7$ cases
and have found good matches with the hypothesis \eq{eq:hypothesis}.
Assuming the hypothesis and reminding the form \eq{eq:freelam}, we obtain  
\[
F_{N,R}(\lambda,k)= \tilde U_d^0 \log (k/\lambda) + \frac{NR}{6} \log \lambda
+\tilde p_{N,R}(k/\lambda) \ \  \hbox{ for } k/\lambda \sim +0 \hbox{ and } R>R_c,
\label{eq:freehigh}
\]
where $\tilde U_d^0:=\lim_{k\rightarrow +0} k \, \langle U_d(\phi) \rangle$,
$\tilde p_{N,R}(k/\lambda)$ is smaller than $\log(k/\lambda)$ in order, and 
\[
\tilde U_d^0 =\frac{NR}{6}-\frac{\# P}{2}+\delta \tilde U_d^0
\]
with $\delta \tilde U_d^0$ sub-leading in large $R$.
Note that $\tilde U_d^0\geq 0$ due to $\langle U_d(\phi) \rangle>0$, 
and therefore $\delta \tilde U_d^0$ takes positive values in the range 
$R_c < R <(N+1)(N+2)/2$.

Here it is a non-trivial question whether $\tilde p_{N,R}(k/\lambda)$ has a finite limit $\tilde p_{N,R}(+0)$.
For example, a slow correction of order $\sim 1/\log(k)$ to $k \langle U_d(\phi) \rangle$ 
for $k\sim +0$ leads to a double logarithmic divergence of $\tilde p_{N,R}(+0)$. However, as shown in Figure~\ref{fig:udkdep}, the data points of $k \langle U_d(\phi) \rangle$ can be fitted very well with 
a correction of order $\sqrt{k}$, and there is no good motivation for introducing such slow corrections.
Therefore it would be reasonable to assume $\tilde p_{N,R}(+0)$ to exist as a finite value.
We also comment that the hypothesis \eq{eq:hypothesis} is nothing but what can be obtained from 
the perturbative computation in the leading order \cite{Lionni:2019rty}, 
as the coincidence in the left figure of Figure~\ref{fig:od} shows.
Therefore $\delta \tilde U_d^0$ is a correction beyond the leading order perturbative
computation.

The difference between the behavior of the free energy \eq{eq:freelow} and \eq{eq:freehigh} 
characterizes the two phases separated by $R=R_c$.
These formulas will be used in Section~\ref{sec:normalizability}, where we will discuss 
the normalizability of the wave function of the tensor model. 

\subsection{Geometric properties}
\label{sec:geometry}
In Section~\ref{sec:comparison}, we have found the deviation between the results of the simulations and the 
perturbative analytic results in the vicinity of the phase transition point. 
This suggests that some non-perturbative configurations are important 
in the vicinity of the phase transition point.
In this subsection, to discuss the characteristics of the configurations around the phase 
transition point, we study the distributions
of the vectors $\phi_a^i\ (i=1,2,\ldots,R)$ in the $N$-dimensional vector space associated to the 
lower index. These vectors define a point cloud with $R$ points, 
where $\phi^i_a$ for each $i$ determines the location
of each point in the $N$-dimensional vector space.  
We study the dimensions of such point clouds.
It turns out that the dimensions depend on the parameters of the matrix model.
In particular, the dimensions take the smallest values 
at the transition point as functions of $R$.

To study the dimension of a point cloud we use angle distributions among the vectors.   
The angle between two vectors, say $\phi_a^i$ and $\phi_a^j$\ $(i\neq j)$, in the $N$-dimensional
vector space is given by
\[
{\rm ang} (\phi^i,\phi^j):=\arccos \left( \frac{\phi_a^i \phi_a^j}{\sqrt{\phi_a^i\phi_a^i \phi_b^j \phi_b^j}}\right).
\] 
Assuming that the vectors approximately form a rotationally symmetric $d$-dimensional point cloud, 
the distribution of the angles should be approximately given by
\[
\rho(\theta)\, d\theta={\cal N} \sin^{d-2}(\theta)\, d\theta,
\label{eq:sindim}
\]
where $\theta$ designates the angle, and ${\cal N}$ is a normalization factor. 
This formula can easily be obtained by radially
projecting points to the unit sphere $S^{d-1}$, and computing infinitesimal areas associated with 
given mutual angles.
The dimensions can be computed by fitting the formula \eq{eq:sindim} to the angle 
distributions obtained from the datas.

\begin{figure}[]
\begin{center}
\includegraphics[clip,width=7.0cm]{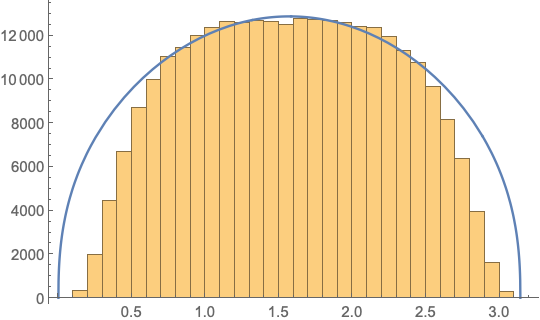}
\hfil
\includegraphics[clip,width=7.0cm]{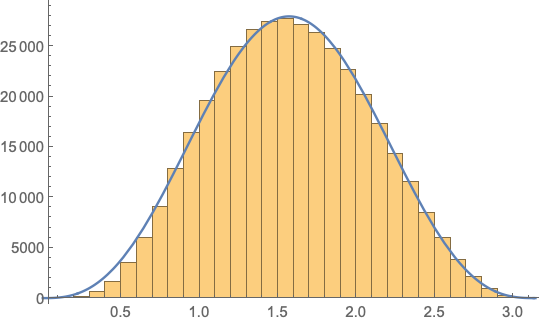}
\caption{
Examples of fitting \eq{eq:sindim} to the histograms of the mutual angles among $\phi^i$s
from the actual datas. The horizontal axes represent the angle $\theta$. 
The parameters are $N=10,\ k=10^{-8}$
with $R=60$ (left) and $R=70$ (right), respectively.  
Fitting is performed only at the 3/5 portion around the center ($\theta\sim \pi/2$),
ignoring 1/5 portions on each side.
The fitted values of dimensions are $d=2.4$ and $d=4.7$, respectively,
in these cases.}
\label{fig:dimexamp}
\end{center}
\end{figure}

\begin{figure}[]
\begin{center}
\includegraphics[clip,width=8.0cm]{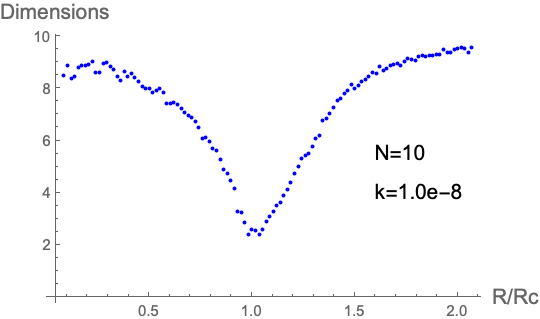}
\caption{
The $R$-dependence of the dimension for $N=10,\ k=10^{-8}$.
It takes the lowest value at the transition point. $R_c=(N+1)(N+2)/2-N+2$ with $N=10$.
Errors are not estimated in this plot.}
\label{fig:dimensions}
\end{center}
\end{figure}

In Figure~\ref{fig:dimexamp}, we show two examples of the fitting. 
As shown in the figures, the fitting is generally quite good for high dimensions
but not so much for lower dimensions.
The reason behind this is that the point clouds cannot be characterised as a single dimensional object but
are a mixture of objects with different dimensions,
as we discuss in Section~\ref{sec:expdimensions}.
Yet, to characterize the configurations in terms of dimensions, we perform the fitting 
restricted to a portion around the center, namely $\theta\sim \pi/2$,
because there exist dominant numbers of cases in this region.
In this sense, the dimension is merely a qualitative characterization, but it still
gives a fairly interesting observable: The dimension takes the lowest value at the phase transition 
point as a function of $R$.
For instance, this can be observed for $N=10,\ k=10^{-8}$ in Figure~\ref{fig:dimensions}. 

\begin{figure}[]
\begin{center}
\includegraphics[clip,width=7.0cm]{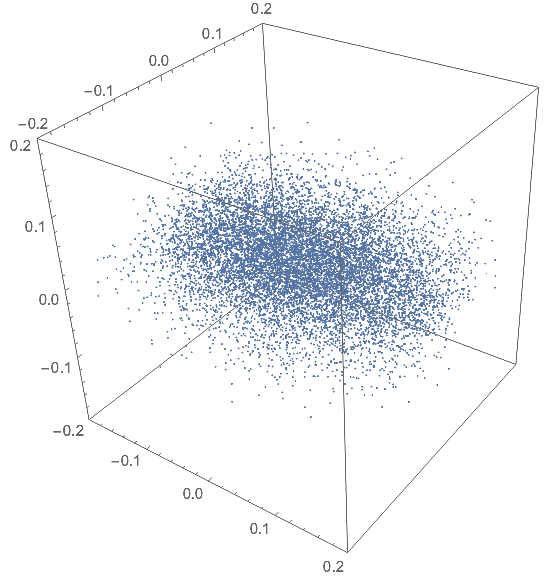}
\hfil
\includegraphics[clip,width=7.0cm]{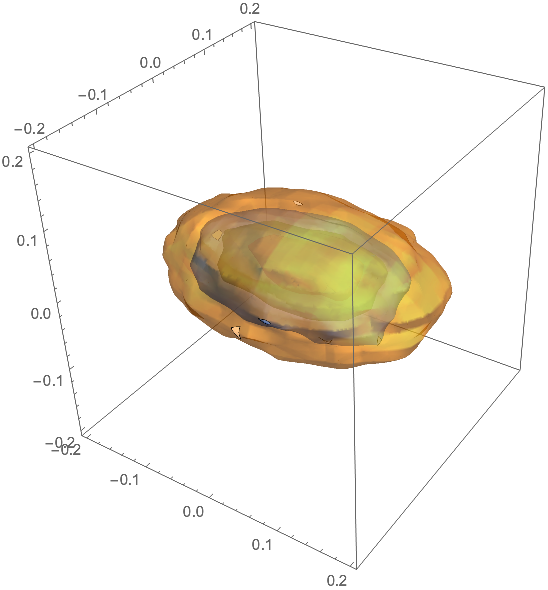}
\caption{
Left: The collection of the point clouds obtained from the simulation with 
$N=10,\ R=57$, and $k=10^{-8}$.
The point cloud from each data of $\phi_a^i$ is projected into the three-dimensional space and the collection
through all the datas are plotted.
For the projection, PCA is used to take three major directions out of $N$ dimensions.
Right: The corresponding density plot.
The shape is like a squashed rugby ball, which may be regarded as an object with 
a dimension between 2 and 3. }
\label{fig:object}
\end{center}
\end{figure} 

 \begin{figure}[]
\begin{center}
\includegraphics[clip,width=10.0cm]{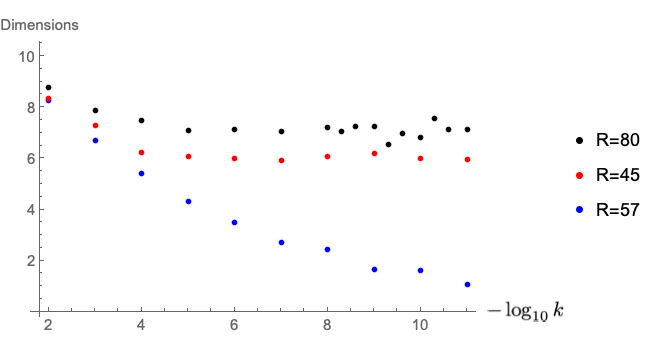}
\caption{The $k$ dependence of the dimensions of the configurations from the data of $N=10$ and 
$R$ shown in the figure. The datas for $R=80$ do not converge well for small $k$ due to the difficulty 
of the simulations explained in Section~\ref{sec:slowdown}.
Errors are not estimated in this plot.}
\label{fig:dimk}
\end{center}
\end{figure} 

It is instructive to directly see a point cloud itself. 
A point cloud exists in an $N$-dimensional space, but if its dynamical dimension is
lower than three, one can project it into a three-dimensional space by extracting the main
three extending directions through principal component analysis (PCA).
Figure~\ref{fig:object} shows a collection of 
a number of such projected point clouds,
which have been sampled from the simulation with $N=10,\ R=57,\ k=10^{-8}$.
According to Figure~\ref{fig:dimensions}, the point cloud has a dimension nearly two
in  this case, and we indeed find an approximately two-dimensional object 
which has the shape of a squashed rugby ball 
as shown in the right figure of Figure~\ref{fig:object}.

Finally, let us discuss the $k$-dependence of the dimension. The general behavior is that the 
dimensions decrease with the decrease of $k$ and converge to limiting values, as is shown in 
Figure~\ref{fig:dimk}.

\subsection{Symmetry breaking}
\label{sec:symmetry}
As shown in Section~\ref{sec:phasetrans}, the phase at $R>R_c$ is characterized by large values of
$\langle r^2 \rangle$.
Since a non-vanishing value of $\phi_a^i$ breaks the $O(N)$ symmetry associated to the 
lower index vector space, the phase at $R>R_c$ will be characterized by symmetry breaking. 
In this subsection, we will study this aspect.

Let us consider one of the generators $T_{ab}$ of $SO(N)$.  
The size of the breaking of $T_{ab}$ by a vector $\phi_a^i$
will be characterized by the size of the vector $T_{ab} \phi_b^i$.
By considering its square and summing over all the vectors, 
the breaking by a configuration can be characterized by
$\sum_{i=1}^R  T_{ab}  T_{ab'} \phi_b^i \phi_{b'}^i$. 
Thus the natural quantity to study is
\[
M_{m\,m'}:= T^{(m)}{}_{ab}T^{(m')}{}_{ab'} \sum_{i=1}^R\phi_b^i \phi_{b'}^i,
\label{eq:defofM}
\] 
where $T^{(m)}{}_{ab}\ (m=1,2,\ldots,N(N-1)/2)$ are a basis of the $so(N)$ generators
with the normalization $T^{(m)}{}_{ab} T^{(m')}{}_{ab}=2 \delta_{mm'}$ for later convenience.  
Note that the definition of $M_{m\,m'}$ conserves the $S_R$ symmetry for the upper index.

An $O(N)$-invariant observable which can be obtained from $M_{m\,m'}$ is the set of the eigenvalues 
of the matrix $M$. 
For an arbitrary $\phi_a^i$, we can diagonalize $m_{ab}:=\sum_{i=1}^R\phi_a^i \phi_{b}^i$ by an $SO(N)$ 
transformation. 
Then it is straightforward to prove that the eigenvalues of $M$ are given by\footnote{This can be proven by
explicitly taking the basis, $T^{(ij)}{}_{ab}:=\delta_{ia}\delta_{jb}-\delta_{ib}\delta_{ja}$.}
\[
eg(M)=\{ e^{\phi}_a + e^{\phi}_{b} \, | \, a,b=1,2,\ldots,N, \ a<b \},
\label{eq:egm}
\]
where $e^{\phi}_a\ (a=1,2,\ldots,N)$ are the eigenvalues of the matrix 
$m$.

\begin{figure}[]
\begin{center}
\includegraphics[clip,width=7.0cm]{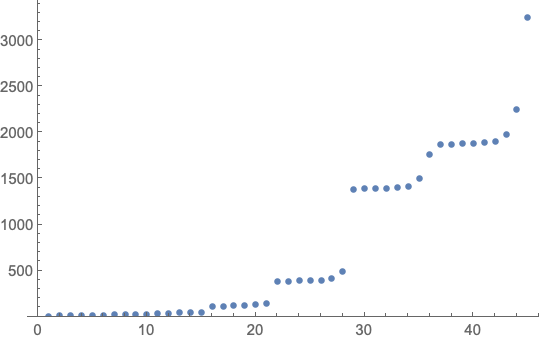}
\hfil
\includegraphics[clip,width=7.0cm]{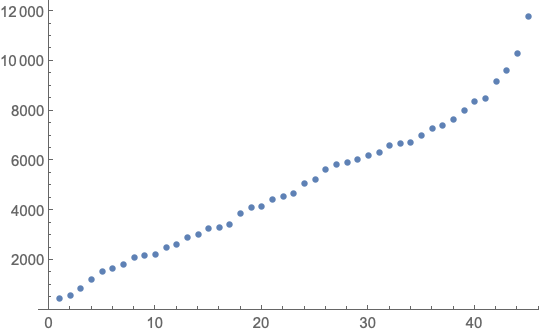}
\caption{$eg(M)$ defined in \eq{eq:egm} are plotted in ascending order for 
a sample of $\phi_a^i$, each from a simulation for $R=60$ (left) and for $R=80$ (right), respectively, 
with $N=10,\ k=10^{-8}$. 
The stair-like pattern in the left figure implies that the 
$SO(N)$ symmetry is hierarchically broken to $SO(N-1),\ SO(N-2),\ldots$.  
In fact, the horizontal locations of the steps 
agree with the numbers of the generators of $SO(n)\ (n=2,3,\ldots,N)$. 
All the symmetries are broken with no hierarchal structure in the right figure.}
\label{fig:stairs}
\end{center}
\end{figure} 

Figure~\ref{fig:stairs} gives two examples of the eigenvalues $eg(M)$. In the figures, the eigenvalues 
are plotted in ascending order along the horizontal direction.
In the case of the left figure, one can find an interesting stair-like pattern of the eigenvalues. 
This  pattern means that the original $SO(N)$ symmetry is hierarchically broken to $SO(N-1),\ SO(N-2),\ldots$.
In fact, the horizontal locations of the steps agree with the numbers of the generators of these symmetries. 
On the other hand, in the case of the right figure, all the symmetries are broken with no obvious
hierarchal structure.

\begin{figure}[]
\begin{center}
\includegraphics[clip,width=7.0cm]{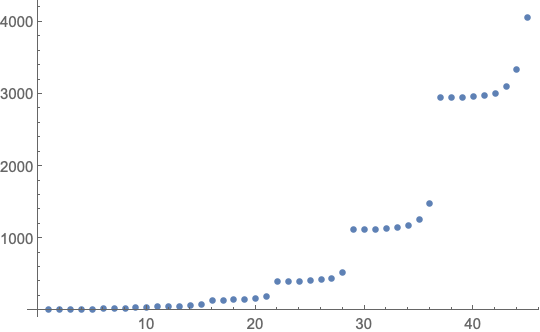}
\hfil
\includegraphics[clip,width=7.0cm]{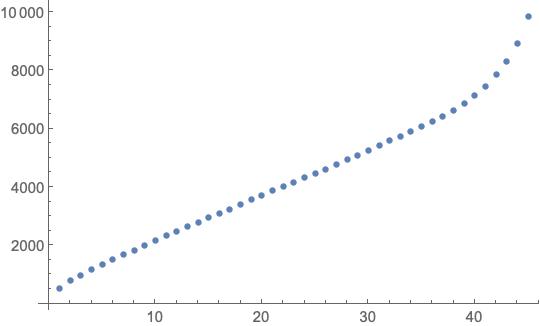}
\caption{The mean values $\langle eg(M)\rangle$ are plotted for the datas of $R=60$ (left) 
and of $R=80$ (right), respectively, with $N=10,\ k=10^{-8}$. }
\label{fig:averagestair}
\end{center}
\end{figure} 
 
Since the pattern above generally fluctuates over the samples of $\phi_a^i$ in a simulation, 
we consider an average, $\langle eg(M) \rangle$. 
The precise definition of this quantity is as follows: we run a simulation with a certain choice of parameters;
for each sample of $\phi_a^i$ in a simulation, we compute eigenvalues $eg(M)$ 
and order them in ascending order; 
then we take mean values of each entry over all the datas of the simulation.
Figure~\ref{fig:averagestair} shows $\langle eg(M) \rangle$ computed from the simulations 
respectively for $R=60$ (left) and for $R=80$ (right) with $N=10,\ k=10^{-8}$.

\begin{figure}[]
\begin{center}
\includegraphics[clip,width=7.0cm]{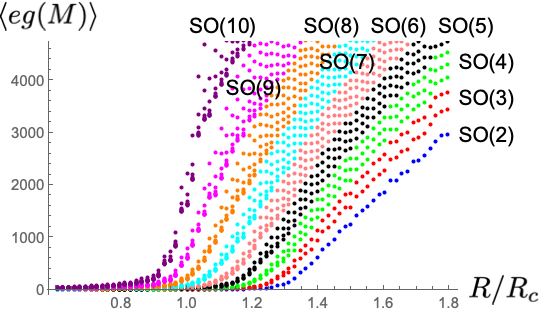}
\caption{The mean eigenvalues $\langle eg(M) \rangle$ against $R/R_c$ 
for $N=10,\ k=10^{-8}$. $R_c=(N+1)(N+2)/2-N+2$. 
The eigenvalues are plotted vertically at each $R$. 
For clear distinction, the points are colored
according to the numbers of the generators of the symmetries.
Each symmetry is broken when the corresponding eigenvalues leave from the horizontal axis. }
\label{fig:breaking}
\end{center}
\end{figure} 
 
Figure~\ref{fig:breaking} shows the dependence of $\langle eg(M) \rangle$ over the change of $R$ 
for $N=10,\ k=10^{-8}$. The eigenvalues start to increase from $R\sim 0.9 R_c$ with the increase of $R$.
The symmetry breaking occurs one by one: first $SO(10)$, then $SO(9)$,
and so on, until finally all the symmetries are broken at $R\sim 1.3 R_c$. 
In the figure, one can find some gaps between
the eigenvalues in the vicinity of $R\sim R_c$. They correspond to the differences of the step heights, 
which for example exist in left figure of Figure~\ref{fig:averagestair}.  
As $R$ becomes larger, the gaps gradually disappear,
approaching the situation in the right figure of Figure~\ref{fig:averagestair}.  

The above symmetry breaking in a cascade manner is consistent with the results in the previous subsections.
When $R<R_c$, since $\langle \phi^2 \rangle$ is small, there is no symmetry breaking. 
As $R$ increases from $R\sim R_c$, the vectors $\phi_a^i \ (i=1,2,\ldots,R)$ start to 
take larger values and fill a subspace, the dimension of which increases with the increase of $R$.  
Since the subspace breaks part of the $SO(N)$ symmetry, depending on its dimensions, 
more symmetries are broken with the increase of $R$.

Let us comment about the fate of the discrete symmetry $\phi_a^i\rightarrow -\phi_a^i\ \forall i$. 
For $N=\hbox{odd}$,
this corresponds to the $Z_2$ subgroup of the $O(N)$ symmetry. The 
quantity\footnote{To balance the normalization with that of $T^{(m)}{}_{ab}$, we put a factor of $1/R$.},  
$\sum_{i=1}^R \phi_a^i/R$, is not invariant 
under the discrete symmetry, and therefore the expectation value of its square, 
$\langle  \sum_{i,j=1}^R \phi_a^i \phi_a^j \rangle/R^2$, will provide a good quantity to measure its breaking.
It has turned out that the expectation values computed from the simulation datas stay small 
in the order $\lesssim O(1)$ over the range, and we have not observed any signatures of its breaking.

\subsection{Slowdown of Monte Carlo updates}
\label{sec:slowdown}
In the previous work \cite{Sasakura:2019hql}, 
we encountered a rather serious difficulty of the Monte Carlo simulation:
For $R\gtrsim R_c$ and small $k$, the step sizes of the simulations had to be tuned very small for
reasonable acceptance rates of Metropolis updates, but then the updates of configurations were
too slow for the system to reach thermodynamic equilibriums within our runtimes.
Therefore, in this paper, we have improved the strategy: Integrating out the radial direction of the model
and using the so-called Hamiltonian Monte Carlo method for simulations. 
Indeed the new strategy drastically improves the efficiency of the simulations, but 
we still encounter the slowdown for smaller $k$, which is however several orders of magnitude smaller than 
that in the previous work. 
This implies that this slowdown is an intrinsic property of the model, which is independent from 
methods of simulations, and would even suggest
a possibility of the presence of a transition to a new phase characterized by slow dynamics.
However, in this subsection, we will show that the last possibility is unlikely,
and the system in the phase at $R>R_c$ is rather like a fluid with a viscosity 
which continuously grows for smaller $k$.

The speed of updates can be quantified by the mean value of distances between 
neighboring configurations in a sequence 
of updates, $\tilde \phi_a^i(1),\ \tilde \phi_a^i(2),\ldots,\ \tilde \phi_a^i(M+1)$:
\[
\langle (\delta \tilde \phi)^2 \rangle := \frac1{M}\sum_{m=1}^{M} \left| \tilde \phi(m+1)-\tilde \phi(m)
\right|^2,
\label{eq:speed}
\] 
where $|X|^2:=\sum_{i=1}^R X_a^iX_a^i$.  
In the ideal maximum situation that each entry of the sequence is independent from the others, 
$\langle (\delta \tilde \phi)^2 \rangle=2$ because of the normalization $|\tilde \phi(m)|^2=1$.    

\begin{figure}[]
\begin{center}
\includegraphics[clip,width=7.0cm]{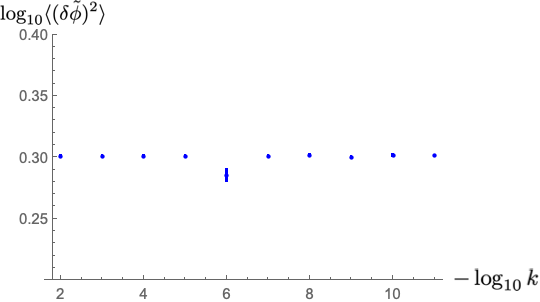}
\hfil
\includegraphics[clip,width=7.0cm]{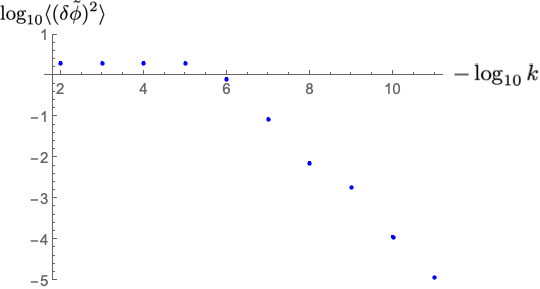}
\caption{
The average speed of updates, $\langle (\delta \tilde \phi)^2 \rangle$, is plotted 
for $R=45$ (left) and $R=80$, respectively, with $N=10$. 
In the simulations, the ideal maximum situation ($\log_{10}2\sim 0.3$) is realized for $R=45$,
while there is a rapid decrease for $R=80$ with the decrease of $k$ for $k\lesssim 10^{-6}$.}
\label{fig:delphi}
\end{center}
\end{figure} 

Figure~\ref{fig:delphi} shows the dependence of $\langle (\delta \tilde \phi)^2 \rangle$ against 
the value of $k$ for $R=45$ (left) and $R=80$ (right), respectively, with $N=10$.
In the simulations, the step sizes, namely the value $\epsilon$ in Section~\ref{sec:MonteCarlo}, 
are properly chosen for reasonable acceptance 
rates\footnote{The acceptance 
rates are typically around from 80 to 99 percents in our simulations.} 
for each $k$, while the other parameters of simulations, such as leapfrog numbers, are fixed. 
The $R=45$ case keeps the ideal values around $\log_{10} 2 \sim 0.3$ throughout the 
shown range of $k$. 
On the other hand, the $R=80$ case has a rapid decrease of the speed with the decrease of $k$
at $k\lesssim 10^{-6}$.
 
\begin{figure}[]
\begin{center}
\includegraphics[clip,width=6.0cm]{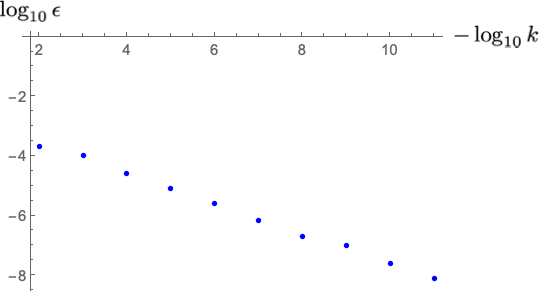}
\hfil
\includegraphics[clip,width=7.0cm]{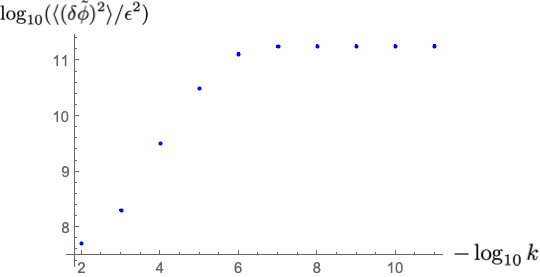}
\caption{Left: The values of $\epsilon$ for the simulation of $N=10,\ R=80$.
Right: The rescaled speed of updates. }
\label{fig:delphieps}
\end{center}
\end{figure} 
 
The speed of updates defined above is dependent on the parameters of the simulation such as step size, 
leapfrog number, and even the frequency at which the data is saved, and is therefore not a quantity intrinsic 
to the model. For instance, the starting point of decreasing, $k\sim 10^{-6}$, has no physical meaning,  
since this can easily be changed by taking different simulation parameters.
However, in the data above, 
the only parameter which is varied is the step size $\epsilon$ among different values of $k$,
and it is therefore meaningful to compare the data for different values of $k$ by 
rescaling $\langle (\delta \tilde \phi)^2 \rangle/\epsilon^2$ to cancel the obvious dependence on $\epsilon$.
The left figure of Figure~\ref{fig:delphieps} plots the values of $\epsilon$ taken for the simulation
of $N=10,\ R=80$, and the right is for the corrected values, $\langle (\delta \tilde \phi)^2 \rangle/\epsilon^2$.
In the right figure, leaving aside the irrelevant ideal region $k\gtrsim 10^{-6}$,  
one can see that the values are almost flat in the region $k\lesssim 10^{-6}$ with no essential change.
This implies that the system is basically similar up to the obvious rescaling among different values of $k$.

We also used parallel tempering \cite{parallel} in addition to the Hamiltonian Monte Carlo method for
taking some datas which systematically study $k$-dependencies. 
The exchanges of configurations were performed among different values of $k$,
typically taken $k=10^{-n}\ (n=2,3,\ldots,11)$, with common values of the other parameters. 
In the region $R>R_c$, as $R$ increases from $R_c$, the exchange rate quickly reduces for the above
choices of $k$'s. Therefore, parallel tempering does not seem effective to solve the slow update problem, 
which exists at $R\gtrsim R_c$ for small $k$.
On the other hand, the exchange rate is high for small $k$ at $R\lesssim R_c$, which can easily be understood
by the presence of the well-defined $k/\lambda \rightarrow +0$ limit at $R<R_c$, as discussed in 
Section~\ref{sec:k0limit}: the sets of configurations are similar among different values of $k$, 
when $k$ is small enough.
However, we did not observe any major differences between the datas with or without parallel tempering.
This would imply that there are no major isolated dominant configurations which can only be reached
by employing parallel tempering. 
All in all, we have not observed any essential improvement by employing parallel tempering 
in addition to the Hamiltonian Monte Carlo method.

\begin{figure}[]
\begin{center}
\includegraphics[clip,width=12.0cm]{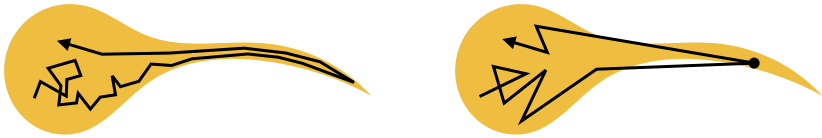}
\caption{Left: Smooth sampling with relatively smaller $\epsilon$. Sampling smoothly 
visit both of the broad and narrow regions. 
Right: Sampling with relatively larger $\epsilon$. 
Sampling mainly moves within the broad region, but is occasionally trapped 
for a while in the narrow region. }
\label{fig:updates}
\end{center}
\end{figure} 

Another interesting aspect of our actual Hamiltonian Monte Carlo simulation is that 
a relatively smaller choice of the step size $\epsilon$
seems to give better sampling, and we even took such small values that acceptance rates were nearly 1.  
This seems to be in contradiction with the more common situation that larger $\epsilon$ with 
a reasonable acceptance rate like several 10\% would give better sampling. 
However, this apparent contradictory aspect could be explained in the following manner in our case. 
The positive semi-definite \eq{eq:positiveU} of the first term in the exponent of
the matrix model \eq{eq:partition} implies that the dominant configurations for small $k$ are around 
$\sum_{i=1}^R \phi_a^i \phi_b^i \phi_c^i\sim 0$, and this condition becomes tighter as 
$\phi_a^i$ can take larger values when $k$ is taken smaller.  
Therefore, the space of dominant configurations can be illustrated as in Figure~\ref{fig:updates}: 
the dominant configuration space is broad in the 
small $\phi_a^i$ region, but it becomes narrower as $\phi_a^i$ becomes larger. Here, we also 
assume that dominant configurations are connected, as suggested in the previous
paragraph. Assuming the dominant configuration space as shown in the figure, 
the updates with relatively smaller $\epsilon$ will smoothly visit the narrow region as well as the 
broad region.
On the other hand, sampling with relatively larger $\epsilon$ mainly moves within the broad region,
occasionally jumps to the narrow region, and is trapped for a while to compensate the low 
possibility to visit the narrow region. We have actually observed such trapping to occur more 
frequently for relatively larger values of $\epsilon$. 
This occasional trapping damages quality of sampling and it generally takes longer time  
to obtain a dataset with lower margins of error.

Let us summarize this subsection.
We encountered the slowdown of the Monte Carlo updates in the region $R\gtrsim R_c$ with small
$k$. The speed of updates becomes slower as $k$ becomes smaller, but the dependence is continuous
and is subject to the explanation with obvious rescaling.
Therefore we have not observed any qualitative changes of the system under the change 
of the value of $k$, and it is unlikely that there is a phase transition to a new phase with 
characteristics of slow dynamics. 
Rather it seems that the system continues to behave like a fluid with a viscosity which 
continuously grows for smaller $k$ in the region $R\gtrsim R_c$.  

\section{Implications to the tensor model}
\label{sec:tensor}
In this section, we discuss the implications of the results of the simulations to 
the tensor model in the canonical formalism, the canonical tensor model~\cite{Sasakura:2011sq,Sasakura:2012fb}.

\subsection{Phase transition point and the consistency of the tensor model}
\label{sec:critical}
The wave function of the canonical tensor model, that is obtained by solving a number of first-class constraints
to the wave function,\footnote{The equations are given by the physical state conditions, $\hat {\cal H}_a|\Psi\rangle=0$ and $\hat{\cal J}_{ab}|\Psi\rangle=0$, where $\hat {\cal H}_a$ and $\hat {\cal J}_{ab}$ are 
the quantized first-class constraints of the tensor model.
See an appendix of \cite{Sasakura:2019hql} for more thorough compact explanations.}
has the following form \cite{Narain:2014cya}, 
\[
\Psi(P)=\left(\int_{\mathbb{R}^N} \prod_{a=1}^N d\phi_a\, \exp \left( I\, P_{abc} \phi_a \phi_b \phi_c
\right) {\rm Ai}\left(\kappa\,\phi_a \phi_a\right) \right)^{\frac{\lambda_H}{2}},
\label{eq:wavefn}
\]
where $P_{abc}$, a real symmetric tensor, is the configuration variable of the tensor model, 
$\lambda_H=(N+2)(N+3)/2$,
$I$ is the imaginary unit (so $I^2=-1$), ${\rm Ai}(\cdot)$ designates the Airy Ai function, and $\kappa$ is a real 
constant in the tensor model. It is particularly important that $\lambda_H$ is determined 
by the hermiticity condition for the Hamiltonian constraint of the tensor model, and 
therefore must have this particular form depending on $N$. 
Physically, 
the sign of the parameter $\kappa$ is supposed to be opposite to that of 
the cosmological constant,
based on the argument relating the mini-superspace approximation of GR and the tensor model with
$N=1$ \cite{Sasakura:2014gia} (See an appendix of \cite{Sasakura:2019hql} for more details).

The simplest observable for the physical state represented by the wave function \eq{eq:wavefn} would be
given by
\s[
\langle \Psi | e^{-\alpha \hat P_{abc} \hat P_{abc} } | \Psi \rangle&= 
\int_{\mathbb{R}^{\# P}} \prod_{a,b,c=1 \atop a\leq b \leq c}^N dP_{abc}\,
e^{-\alpha P_{abc}P_{abc} } \left| \Psi(P)\right| ^2 \\
&=
{\cal N} \alpha^{-\#P/2} \int_{\mathbb{R}^{NR}} \prod_{a=1}^{N}\prod_{i=1}^R d\phi_{a}^i
\exp\left(
-\frac{1}{4 \alpha} \sum_{i,j=1}^R U_{ij}(\phi) \right) \prod_{i=1}^R {\rm Ai}\left(\kappa\,\phi_a^i \phi_a^i\right),
\label{eq:obstensor}
\s]
where $R=\lambda_H$, we have introduced replicas $\phi_a^i\ (i=1,2,\ldots,\lambda_H)$ 
to replace the power  coming from that of \eq{eq:wavefn} in the first line, and
have performed the Gaussian integration over $P_{abc}$.
Here $\alpha$ is an arbitrary positive number, ${\cal N}$ is an unimportant factor independent from $\alpha$, 
$\# P=N(N+1)(N+2)/6$, i.e. the number of independent components of $P_{abc}$. 

The system \eq{eq:obstensor} is complicated due to the presence of the Airy functions. 
However, 
when $\kappa$ is taken to be positive, which physically corresponds to a negative cosmological constant,
the Airy function ${\rm Ai}\left(\kappa\,\phi_a^i \phi_a^i\right)$ is a function 
that rapidly decays with the increase of $\phi_a^i\phi_a^i$. 
Therefore, as an interesting simplification, we could replace the Airy function by a rapidly damping
function with a simpler form. In particular, to make the correspondence to
the matrix model \eq{eq:partition}, we consider a simplified wave function, 
\[
\Psi_{simple}(P)=\left(\int_{\mathbb{R}^N} \prod_{a=1}^N d\phi_a\, \exp \left( I\, P_{abc} \phi_a \phi_b \phi_c
-k( \phi_a \phi_a)^3 \right) \right)^{\frac{R}{2}}
\label{eq:simplify}
\] 
with $R=\lambda_H$ and a positive $k$ by performing the replacement
${\rm Ai}\left(\kappa\,\phi_a^i \phi_a^i\right)\rightarrow \exp \left( -k (\phi_a^i \phi_a^i)^3\right)$
in \eq{eq:wavefn}. With the observable mentioned above, this leads to
\s[
\langle \Psi| e^{-\alpha \hat P_{abc} \hat P_{abc} } | \Psi \rangle
&\sim\langle \Psi _{simple}| e^{-\alpha \hat P_{abc} \hat P_{abc} } | \Psi_{simple} \rangle\\
&=
\int_{\mathbb{R}^{\# P}} \prod_{a,b,c=1 \atop a\leq b \leq c}^N dP_{abc}\,
e^{-\alpha P_{abc}P_{abc} } \left| 
\Psi_{simple}(P)\right|^{2}\\
&=
{\cal N} \alpha^{-\#P/2} \int_{\mathbb{R}^{NR}} \prod_{a=1}^{N}\prod_{i=1}^{R} d\phi_{a}^i
\exp\left(
-\frac{1}{4 \alpha} \sum_{i,j=1}^{R} U_{ij}(\phi)-k \sum_{i=1}^{R} (\phi_a^i\phi_a^i)^3 \right)
\\
&= {\cal N} \alpha^{-\#P/2} Z_{N,R}
\left(\frac{1}{4\alpha},k\right),
\label{eq:reltensorz}
\s]
where $R=\lambda_H$.

One important matter in the relation \eq{eq:reltensorz} between the 
tensor and matrix models is that the parameter $R$ of the matrix model 
\eq{eq:partition} is related to $N$ by $R=\lambda_H=(N+2)(N+3)/2$.
What is striking is that this value agrees with the critical value $R_c\sim (N+1)(N+2)/2-N+2$ 
in the leading order of $N$. Considering the ambiguity of the approximate relation \eq{eq:reltensorz}, 
we could say that the tensor model is exactly on or at least in the vicinity of (or a little above of) 
the continuous phase transition point of the matrix model.
This is quite intriguing, because our common knowledge tells that continuum theories 
can often be obtained by taking continuum limits around continuous phase transition points 
in discretized theories. 
We could say that 
the consistency of the tensor model automatically puts the tensor model at the location where a continuum limit
may be feasible, though it is currently difficult to conclude this 
because of the ambiguity contained in the simplification above.
 
\subsection{Dimensions and symmetries of the configurations}
\label{sec:expdimensions}
In this subsection we will discuss the results of the simulations concerning 
dimensions and symmetries obtained in Section~\ref{sec:results}.
For this purpose we refer to a property of the wave function \eq{eq:wavefn} 
that the peaks (ridges) of the wave function 
are located on the values of $P_{abc}$ which are invariant under Lie-group transformations.
This symmetry highlighting phenomenon has been found in \cite{Obster:2017dhx,Obster:2017pdq},
where the qualitative argument was given as follows. 
The integration \eq{eq:wavefn} is of an integrand which oscillates rather widely 
due to the pure imaginary cubic function in the exponent. Therefore, for a ``generic" value of $P_{abc}$,
the contributions from different integration spots generally have different phases and mutually cancel among
themselves so that the total amount of integration does not take a large value.
However, at the location where $P_{abc}$ is invariant under a representation $H$ of a Lie group,
$P_{abc}=h_a^{a'}h_b^{b'}h_c^{c'}P_{a'b'c'}\ (\forall h\in H)$, the integration
along the gauge orbit $h_a^{a'} \phi_{a'}\ (\forall h\in H)$ contributes coherently in \eq{eq:wavefn},
and the wave function has the chance to take a large value compared to that at a ``generic" location.
This is indeed realized and has concretely been shown for some tractable cases in \cite{Obster:2017dhx,Obster:2017pdq}.
 
The above qualitative argument will hold at least partially after the simplification \eq{eq:simplify},
since we can expect a similar coherence phenomenon in this case, too.
Then, from the relation \eq{eq:reltensorz}, the symmetry highlighting phenomenon of the wave function 
explained above will 
have a corresponding phenomenon in the matrix model \eq{eq:partition} \cite{Sasakura:2019hql}.
Note that this will be valid for general values of $R$, since 
the constraint $R=\lambda_H=(N+2)(N+3)/2$ coming from the consistency of the tensor model 
has nothing to do with the equalities in \eq{eq:reltensorz}.
In the relation \eq{eq:reltensorz}, the contribution of a peak with a Lie group representation $H$ 
in the second line will correspond on the matrix model side to the contributions of
$N$-dimensional vectors $\phi_a^i \ (i=1,2,\ldots,R)$ being distributed along a gauge orbit
$h_a^{a'} \phi_{a'}\ (\forall h\in H)$. 
In the simulation data, such distributed vectors will appear as a point cloud discussed in 
Section~\ref{sec:geometry}.
This point cloud will have the dimension of the Lie group representation, 
and will break part of the $SO(N)$ symmetry which is not commutative with $H$.
Generally, the wave function contains a number of peaks with various Lie group representations,
and therefore the point cloud will be that of a mixture of various gauge orbits.
This mixed structure will induce a non-obvious pattern of symmetry breaking,
which would be consistent with the hierarchical symmetry breaking in Section~\ref{sec:symmetry}. 

\begin{figure}[] 
\begin{center}
\includegraphics[width=10.0cm]{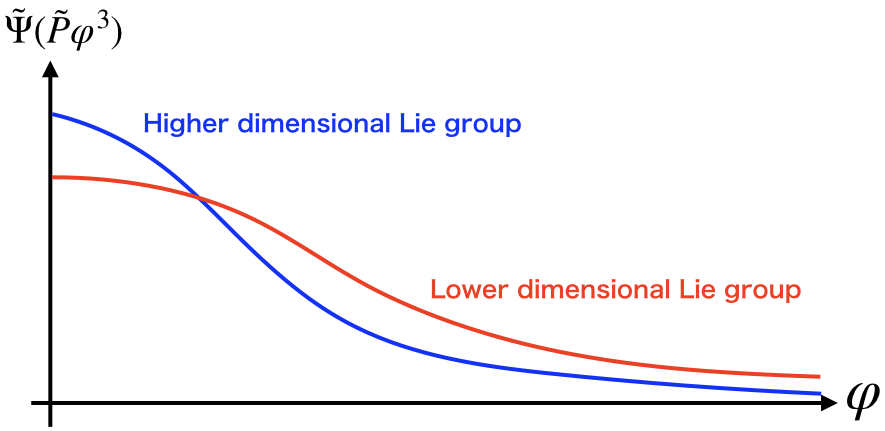}
\caption{A possible profile of $\tilde \Psi(\tilde P \varphi^3 )$,
depending on the Lie-group symmetry of $\tilde P_{abc}$. }
\label{fig:profile}
\end{center}
\end{figure} 

To get more information from the behavior obtained in Section~\ref{sec:results}, let us 
rewrite the second line in \eq{eq:reltensorz} as
\s[
&\int_{\mathbb{R}^{\# P}} \prod_{a,b,c=1 \atop a\leq b \leq c}^N dP_{abc}\,
e^{-\alpha P_{abc}P_{abc} } \left(\int_{\mathbb{R}^N} \prod_{a=1}^N d\phi_a\, \exp \left( I\, P_{abc} \phi_a \phi_b \phi_c
-k( \phi_a \phi_a)^3 \right) \right)^{R}
\\
&={\cal N} \int_0^\infty dP  P^{\#P-NR/3-1} e^{-\alpha P^2} 
\int_{S^{\#P-1}} d\tilde P  
 \left(\int_{\mathbb{R}^N} \prod_{a=1}^N d\varphi_a\, \exp \left( I\, \tilde P_{abc} \varphi_a \varphi_b \varphi_c
-k( \varphi_a \varphi_a)^3/P^2 \right) \right)^{R}  \\
&=
{\cal N'} \int_0^\infty dP  P^{\#P-NR/3-1} e^{-\alpha k P^2} 
\int_{S^{\#P-1}} d\tilde P  
 \left(\int_{0}^\infty d\varphi \,\varphi^{N-1}\ \tilde\Psi\left(\tilde P \varphi^3\right) e^{-\varphi^6/P^2} \right)^{R},
\label{eq:bytildep}
\s]
where ${\cal N},\ {\cal N'}$ are some unimportant coefficients.
Here, to the second line, we have separated $P_{abc}$ into the radial and angular variables, 
$P_{abc}=P \tilde P_{abc}$,
where $P=\sqrt{P_{abc}P_{abc}}$, and have introduced a rescaled variable, $\varphi_a=P^{1/3} \phi_a$.
Then, to the last line, we have rescaled $P^2\rightarrow k P^2$, have divided $\varphi_a$ into
the radial and angular variables, $\varphi_a=\varphi \tilde \varphi_a$ with $\varphi=\sqrt{\varphi_a \varphi_a}$,
and have introduced 
\[
\tilde\Psi\left(\tilde P \varphi^3\right) :=\int_{S^{N-1}} d\tilde \varphi \, e^{I\tilde P_{abc} \varphi^3 \tilde\varphi_a
\tilde\varphi_b\tilde\varphi_c}.
\label{eq:tildepsi}
\] 

The function $\tilde\Psi(\tilde P \varphi^3)$ will have a number of peaks 
at Lie-group symmetric $\tilde P_{abc}$. On such a peak, the value of 
$\tilde\Psi(\tilde P \varphi^3)$ will generally become smaller as $\varphi$ increases, because
the oscillation of the integrand in \eq{eq:tildepsi} will become wilder.  
In the following paragraphs
we will further argue that the $\varphi$-dependence of $\tilde\Psi(\tilde P \varphi^3)$ 
qualitatively depends on the symmetry of $\tilde P_{abc}$ as in Figure~\ref{fig:profile}
to explain the dimensional behavior in Figure~\ref{fig:dimensions}: Namely,
for $\tilde P_{abc}$ symmetric under higher dimensional Lie-groups, 
$\tilde\Psi(\tilde P \varphi^3)$ takes larger values at small $\varphi$ 
but quickly decays with $\varphi$, 
while it takes smaller values at small $\varphi$ but slowly decays with $\varphi$ 
for $\tilde P_{abc}$ symmetric under lower dimensional Lie-groups. 

To see how the dimensional behavior in Figure~\ref{fig:dimensions} can be explained by the profile in Figure~\ref{fig:profile}, 
let us first consider $R<(N+1)(N+2)/2$. 
In this case, the power of $P$ in the integrand of the last line of \eq{eq:bytildep} is positive,
and therefore the integral over $P$ will be over the range $0\leq P \lesssim 1/\sqrt{\alpha k}$ 
with some preference to larger $P$. 
As $k$ is taken smaller,  the larger region of $\varphi$ in the integral of \eq{eq:bytildep}
becomes more dominant, making the peaks associated with lower dimensional Lie-groups
more dominant than higher dimensional ones. Then the increase of the power $R$ in \eq{eq:bytildep} 
will enhance the peaks of lower dimensional Lie groups.
This explains the decrease of the dimensions with the increase of $R$ in the region $R<R_c$ 
in Figure~\ref{fig:dimensions}. 

Let us next consider $R>(N+1)(N+2)/2$. In this case the power of $P$ in the integrand of \eq{eq:bytildep}
is negative, and $P$ will have the preference to smaller values as $R$ increases. 
Then, the last term in \eq{eq:bytildep} will bound the range of $\varphi$ 
in the integration, as $R$ increases.
Because of the profile in Figure~\ref{fig:profile}, increase of $R$ will enhance the peaks with
higher dimensional Lie groups, explaining the increase of dimensions in the region $R>R_c$ 
in Figure~\ref{fig:dimensions}.

\subsection{Normalizability of the wave function of the tensor model}
\label{sec:normalizability}
From the physical point of view, it would be interesting to discuss the norm of the wave function
of the tensor model. 
If the wave function of the tensor model successfully represents a spacetime in some manner, 
the norm of the wave function will linearly diverge in the time direction, which 
is supposed to form a trajectory in the space of $P_{abc}$. 
Thus, the normalizability of the wave function has a connection to the question concerning time in the 
tensor model.   
   
As an approximation or as an example case study similar 
to the actual case, we discuss the norm of the simplified wave function \eq{eq:simplify}. 
More precisely, we study the $\alpha\rightarrow +0$ limit of the relation \eq{eq:reltensorz}:
\[
\lim_{\alpha \rightarrow +0}
\int_{\mathbb{R}^{\# P}} \prod_{a,b,c=1 \atop a\leq b \leq c}^N dP_{abc}\,
e^{-\alpha P_{abc}P_{abc} } \left| 
\Psi_{simple}(P)\right|^{2}=
\lim_{\alpha \rightarrow +0}
{\cal N} \alpha^{-\#P/2} Z_{N,R}\left(\frac{1}{4\alpha},k\right),
\label{eq:relnormz}
\]
where ${\cal N}$ is an unimportant factor independent of $\alpha$.

When $R<R_c$, by putting \eq{eq:freelow} with $\lambda=1/(4 \alpha)$ on the righthand side
of \eq{eq:relnormz}, 
the dominant $\alpha$ dependence in the $\alpha \rightarrow +0$ limit is obtained as 
\[
\alpha^{-\#P/2} Z_{N,R}\left(\frac{1}{4\alpha},k\right) \sim  \alpha^{NR/6-\# P/2}.
\]
This concludes that the norm diverges for this case by assuming 
that the critical value satisfies $R_c<(N+1)(N+2)/2$.

On the other hand, when $R>R_c$, by putting \eq{eq:freehigh} with $\lambda=1/(4 \alpha)$
on the righthand side, we obtain
\[
\alpha^{-\#P/2} Z_{N,R}\left(\frac{1}{4\alpha},k\right) \sim \alpha^{-\delta \tilde U_d^0},
\label{eq:limitabove}
\]
under the assumption that $\tilde p(+0)$ is finite, which has been supported from the data  
in Section~\ref{sec:k0limit}.
This is divergent in the limit $\alpha\rightarrow +0$ in the range $R_c<R<(N+1)(N+2)/2$,
since $\delta \tilde U_d^0>0$, as discussed in Section~\ref{sec:k0limit}.
On the other hand, since $R=\lambda_H=(N+2)(N+3)/2$ of the tensor model is in the 
region $R >(N+1)(N+2)/2$, we cannot currently determine 
whether the simplified wave function of the tensor model is normalizable or not or how rapidly this diverges
if it does. As explained in Section~\ref{sec:k0limit}, this is beyond the leading order perturbative computation.

The simplification \eq{eq:simplify} of the real wave function 
\eq{eq:wavefn} is to approximate the case with a positive $\kappa$, 
which corresponds to a negative cosmological constant. 
Therefore, it is an interesting future study to determine $\delta \tilde U_d^0$ 
to answer the physical question concerning 
the emergence of time in the tensor model for a negative cosmological constant. 
Note also that the above discussion deals with finite $N$, and therefore taking $N\rightarrow \infty$ 
would also require more study on this matter.
  
\section{Summary and future prospects}
\label{sec:summary}
In this paper, we have numerically studied a matrix model 
with non-pairwise index contractions by Monte Carlo simulations. 
The matrix model has an intimate connection to the canonical tensor model, 
a tensor model for quantum gravity in the Hamilton formalism~\cite{Sasakura:2011sq,Sasakura:2012fb}, and 
also has a similar structure as a matrix model that appears in the replica trick of the spherical $p$-spin 
model ($p=3$) for spin glasses~\cite{pspin,pedestrians}.  
The matrix model had previously been analyzed by a few analytic methods and 
Monte Carlo simulations in~\cite{Lionni:2019rty,Sasakura:2019hql}, 
which had suggested the presence of a continuous phase transition around $R\sim N^2/2$. 
This relation between $N$ and $R$ is particularly interesting, 
because this agrees with a consistency condition of the tensor model in the leading order of $N$, 
implying that the tensor model is automatically located exactly on or near a
continuous phase transition point. 
However, in the previous works the evidence for the phase transition was not very clear.   
In this paper we have presented a new set up for Monte Carlo simulations 
by first integrating out the radial direction, and have studied the model by employing the more efficient 
Hamiltonian Monte Carlo method.
We have obtained considerable improvement of the efficiency of the simulations, 
and have found a rather sharp continuous phase transition around $R=R_c\sim (N+1)(N+2)/2-N+2$.
We have also studied various properties of the phase transition and the two phases:
the dimensions of the configurations take the smallest values at the transition point; 
the phase at $R>R_c$ is characterized by cascade symmetry breaking; and the $k/\lambda 
\rightarrow +0$ limit is convergent in one phase and diverges in the other.

We have also discussed some implications to the tensor model. 
In particular, the most striking is the coincidence above between the location of the continuous phase transition
point and the consistency condition of the tensor model in the leading order of $N$.
A well known fact is that continuum theories can often be obtained by taking a continuum 
limit near a continuous phase transition point.
This means that the tensor model seems to automatically put itself at the location where it is possible to find a sensible continuum limit. 
We have also discussed the wave function of the tensor model by using the connection between 
the matrix model and an approximation of a known exact wave function of the tensor model. 
In particular, we have provided a qualitative argument for the dependence on $R$ of both the dimension of the 
preferred class of configurations and the observed symmetry breaking patterns, 
using the intimate connection between Lie-group symmetries and peaks of the wave function as has been investigated before~\cite{Obster:2017dhx,Obster:2017pdq}.

While we have numerically obtained a rather clear picture of the phase structure of the matrix model,
we are still seriously lacking analytic understanding. As shown in Section~\ref{sec:comparison}, 
there seem to exist essential differences between the numerical results and the analytic perturbative 
results performed in \cite{Lionni:2019rty,Sasakura:2019hql}. 
Moreover, other than the qualitative argument given in Section~\ref{sec:expdimensions},
a more rigorous understanding of the behavior of the dimensions and the symmetry breaking would be desirable. 
Analytic understanding is also necessary to discuss the continuum limit discussed  
in the previous paragraph, since taking a large $N$ limit while simultaneously tuning $R$ and $k$  
is difficult to do exclusively through numerical methods.     
Developing an analytical non-perturbative understanding is an important
future direction to understand the dynamics of the matrix model.

Though this paper has given several clear pieces of evidence for the phase transition
in the matrix model, explaining its interesting connection to the canonical tensor model, 
various things still need to be explored in order to understand more about the canonical tensor model
through matrix models of the similar sort.
Most importantly, the simplification of the wave function discussed in Section~\ref{sec:critical} 
by approximating the Airy function for $\kappa>0$ by a conveniently chosen damping function 
does not explain whether the obtained results are universal under a different choice of a damping function.
Moreover, from a physical point of view
we would like to explore the $\kappa <0$ case corresponding to the positive cosmological constant,
rather than the $\kappa>0$ case corresponding to the negative cosmological constant.
In the case of $\kappa<0$, 
the Airy function becomes oscillatory, and the dynamics will most likely be different from the $\kappa>0$ case.
Since this case suffers from the notorious sign problem, it is technically very challenging.
It would also be an interesting future direction to apply new Monte Carlo methods developed to analyze various
other systems suffering from sign problems to our case, as well as to develop analytical treatment.

\vspace{.3cm}
\section*{Acknowledgements}
The Monte Carlo simulations in this work were mainly carried out on XC40 at YITP in Kyoto University. 
N.S. would like to thank S.~Takeuchi for some discussions at the initial stage of the present work. 
The work of N.S. is supported in part by JSPS KAKENHI Grant No.19K03825. 


\end{document}